\documentclass[aps,pra, twocolumn,superscriptaddress]{revtex4-1}

\usepackage{amssymb}
\usepackage{amsmath}
\usepackage{enumerate}
\usepackage[dvipsnames]{xcolor}
\usepackage{graphicx}	
\usepackage{upgreek}
\newcommand{\nc}{\newcommand}
\nc{\eq}{\begin{equation}}
\nc{\eeq}{\end{equation}}
\nc{\eqa}{\begin{eqnarray}}
\nc{\eeqa}{\end{eqnarray}}
\nc{\ar}{\begin{array}}
\nc{\ear}{\end{array}}
\nc{\bfig}{\begin{figure}}
\nc{\efig}{\end{figure}}
\nc{\dg}{\dagger}
\nc{\sx}{\sigma_x}
\nc{\sy}{\sigma_y}
\nc{\sz}{\sigma_z}
\nc{\spl}{\sigma_+}
\nc{\sm}{\sigma_-}
\nc{\nn}{\nonumber}
\nc{\bl}{\color{blue}}
\nc{\rd}{\color{red}}

\def\ket#1{\mathinner{|{#1}\rangle}}

\begin{document}

\title{Proposal for detection of a single electron spin in a microwave resonator}

\author{P. Haikka}
\email[]{pinja@phys.au.dk}
\affiliation{Department of Physics and Astronomy, Aarhus University, Ny Munkegade 120, DK-8000 Aarhus C, Denmark}

\author{Y. Kubo}
\affiliation{Quantronics Group, SPEC, CEA, CNRS, Universit\'e Paris-Saclay, CEA-Saclay, 91191 Gif-sur-Yvette, France}
\affiliation{Okinawa Institute of Science and Technology (OIST) Graduate University, Onna, Okinawa 904-0495, Japan}

\author{A. Bienfait}
\affiliation{Quantronics Group, SPEC, CEA, CNRS, Universit\'e Paris-Saclay, CEA-Saclay, 91191 Gif-sur-Yvette, France}

\author{P. Bertet}
\affiliation{Quantronics Group, SPEC, CEA, CNRS, Universit\'e Paris-Saclay, CEA-Saclay, 91191 Gif-sur-Yvette, France}

\author{K. M\o lmer}
\email[]{moelmer@phys.au.dk}
\affiliation{Department of Physics and Astronomy, Aarhus University, Ny Munkegade 120, DK-8000 Aarhus C, Denmark}

\date{\today}

\begin{abstract}
We propose a method for detecting the presence of a single spin in a crystal by coupling it to a high-quality factor superconducting planar resonator. By confining the microwave field in a constriction of nanometric dimensions, the coupling constant can be as high as $5-10$\,kHz. This coupling affects the amplitude of the field emitted by the resonator, and the integrated homodyne signal allows detection of a single spin with unit signal-to-noise ratio within few milliseconds. We further show that a stochastic master equation approach and a Bayesian analysis of the full time dependent homodyne signal improves this figure by $\sim 30\%$ for typical parameters.
\end{abstract}

\pacs{}

\maketitle

\section{Introduction}

Because of their long coherence times, spins in solids are attractive candidates for quantum information processing. In pure and nuclear-spin-free crystals, electron spins such as NV centers or donors in silicon can reach second-long coherence times using dynamical decoupling sequences~\cite{Bargill.NatCom.4.1743(2013),tyryshkin_electron_2012,wolfowicz_atomic_2013}. Nuclear spins can reach even longer coherence times, up to several hours as demonstrated recently~\cite{steger_quantum_2012,Saeedi2013,Zhong.Nature.517.177(2015)}. While these experiments were carried out on large ensembles of spins, manipulation, read-out, and entanglement of individual spins remain outstanding challenges. Single-spin readout has been demonstrated by several methods. Spin-to-charge conversion has been used for detecting electron spins in electrostatically-defined quantum dots in two-dimensional electron gases~\cite{Elzerman.Nature.430.431(2004),Veldhorst2014,Petersson.Nature.490.380(2012)} as well as in individual donors~\cite{morello_single-shot_2010,pla_single-atom_2012}, and even the nuclear spin of individual molecular magnets~\cite{vincent2012electronic,Thiele.PhysRevLett.111.037203(2013)}). Spin-dependent photoluminescence has enabled the detection of the spin of individual molecules~\cite{Kohler.Nature.363.242(1993),Wrachtrup.Nature.363.244(1993)}, and defect centers in wide-gap semiconductors such as diamond~\cite{jelezko_observation_2004,Manson.PhysRevB.74.104303(2006)} or silicon carbide~\cite{Widmann.NatureMat.14.164(2015)}). Scanning-probe techniques have also been successfully employed for single-spin detection, either with mechanical resonators~\cite{Rugar.Nature.430.329(2004)}, nitrogen-vacancy magnetometers~\cite{Grinolds.NaturePhys.9.215(2013)}, or scanning tunneling microscope tips\cite{Manassen.PhysRevLett.62.2531(1989),baumann2015electron}.

Here we discuss another method, consisting in pushing the principle of inductive detection~\cite{SchweigerEPR(2001)}, which is the basis of all existing commercial electronic paramagnetic resonance (EPR) spectrometers, to the single-spin limit. Inductive detection of EPR proceeds by inserting a sample that contains the paramagnetic impurities of interest in a microwave resonator of frequency $\omega_r$. In continuous-wave (CW) EPR spectroscopy, it is the microwave absorption that occurs when the spin Larmor frequency tuned by a dc magnetic field $B_0$ matches $\omega_r$ which is detected, giving rise to a dip in the transmitted signal amplitude. The most sensitive spectrometers based on inductive-detection so far are able to detect $\sim 10^6$spins$/\sqrt{\mathrm{Hz}}$~\cite{Sigillito.APL.104.104.22407(2014)}; reaching the single-spin limit requires therefore a gain of several orders of magnitude in sensitivity. A first step in that direction was taken in~\cite{bienfait2015reaching}, where a sensitivity of $2000$ spins$/\sqrt{\mathrm{Hz}}$ was obtained by using a high-quality-factor micron-scale superconducting resonator, combined with a Josephson parametric amplifier~\cite{Castellanos-Beltran.ApplPhysLett.91.083509(2007),bergeal_phase-preserving_2010,Zhou.PhysRevB.89.214517(2014)} to detect a spin-echo signal at the quantum limit of sensitivity (see also~\cite{eichler2016electron}). In order to reach single-spin sensitivity, it is essential to enhance the spin-resonator coupling constant $g$ compared to $g/2\pi = 50$\,Hz as obtained in~\cite{bienfait2015reaching}. One possibility is to hybridize the spin and the charge degree of freedom, as proposed in~\cite{Cottet.PhysRevLett.105.160502,tosi2015silicon,beaudoin2016coupling,Samkharadze.PhysRevApplied.5.044004} and recently demonstrated with a carbon nanotube quantum dot~\cite{Viennot.Science.349.408(2015)}. This comes nevertheless at the expense of a reduced coherence time because of the ubiquitous charge noise. 

Here we propose instead to enhance $g$ by incorporating a nanometric constriction in the resonator, as also proposed in~\cite{Tosi.AIPAdv.4.087122(2014)}, which makes it possible to reach $g/2\pi \sim 5 - 10$\,kHz for realistic resonator design. We predict that the absorption dip in the integrated homodyne signal due to the presence of a single spin should be detectable with a unit signal-to-noise ratio in a detection time of few milliseconds, corresponding to a sensitivity of $\sim 0.1\,$spin$\,/\sqrt{\mathrm{Hz}}$. We also analyze the system from a quantum optics perspective, beyond the simple integration of the homodyne signal, using a quantum trajectory formalism. Transient correlations in the signal carry information about the system beyond the steady state mean values and we show that to discriminate the presence or absence of a spin with a given confidence, the full trajectory analysis allows reducing the measurement time by a further $\sim 30\%$.

The Article is structured as follows: In section~\ref{sec:model}, we present the physical system. In Sec. \ref{sec:framework}, we introduce the master equation for the average system dynamics and we determine the mean amplitude of the signal emitted by the resonator and the fluctuations of its integral over time. In Sec.\ref{sec:bayes}, we introduce the quantum trajectory dynamics of the spin-resonator system, conditioned upon the noisy homodyne signal detection, and we present a Bayesian analysis of the information available from the full homodyne detection record. In the conclusion, Sec. \ref{sec:conclusions}, we summarize our results.

\section{System description/Physical Implementation}
\label{sec:model}
%

\subsection{Proposed Setup}

\begin{figure}[h]
\centering
\includegraphics[width=\hsize]{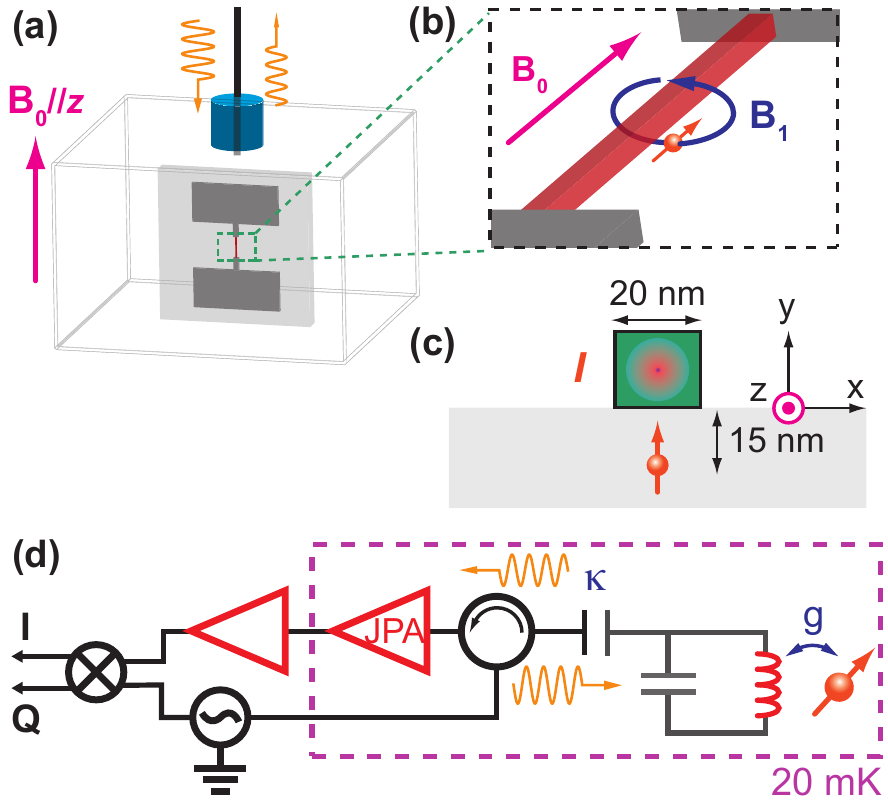}
\caption{Proposed setup.
(a) A superconducting $LC$ resonator consisting of two pads (capacitors $C$) and wire (inductance $L$) is placed in a three dimensional microwave cavity.
(b) A nanometeric scale constriction is made at the center of the inductance wire, below which a single spin is located at 15\,nm.
(c) A cross-section of the structure.
(d) Schematic of the considered measurement circuit.
The signal leaking out of the cavity is first amplified by a quantum limited Josephosn parametric amplifier (JPA), followed by cryogenic low noise HEMT and room temperature amplifiers. }
\label{fig:setup}
\end{figure}

The proposed experimental setup for single spin detection is depicted in Fig.~\ref{fig:setup}. The spin is magnetically coupled to a planar superconducting lumped-element LC resonator, with frequency $\omega_r$. In order to enhance the spin-resonator interaction, the microwave field is strongly confined in the vicinity of the spin with the help of a short (typically $200$\,nm long) superconducting nanowire which is embedded in the middle of the resonator inductance~\cite{Tosi.AIPAdv.4.087122(2014)}. The resonator is probed with microwave signals sent via a capacitively coupled antenna. A possible arrangement, shown in Fig.~\ref{fig:setup}, is to have the sample mounted in a metallic enclosure, and the antenna soldered onto a microwave connector mounted on the enclosure~\cite{bienfait2015reaching}. We assume that the resonator-antenna coupling is chosen such that the loaded quality factor reaches $\sim 10^5$, whereas the internal resonator losses are negligible due to superconductivity; in that regime, all the available signal is emitted into the measurement line, thus maximizing the measurement efficiency. After routing by a circulator, this signal is first amplified by a Josephson parametric amplifier (JPA), then by a low-noise HEMT amplifier at 4K before demodulation at room-temperature, yielding the two signal quadratures ${I(t),Q(t)}$ (see Fig.~\ref{fig:setup}).

A key aspect of the proposal is the use of a JPA to amplify the spin signal. The JPA is an ultra-low-noise microwave amplifier recently developed in the context of circuit QED with performance close to the quantum limit~\cite{Castellanos-Beltran.ApplPhysLett.91.083509(2007),bergeal_phase-preserving_2010,Zhou.PhysRevB.89.214517(2014)}. This performance is conveniently quantified by the quantum efficiency parameter $\eta \equiv 1/(1+N) $, $N$ being the number of noise photons added during the detection process. If the signal detection is performed using the HEMT amplifier with a system noise temperature $T_N \sim 15\,$K, $N= k T_N / (\hbar \omega_r)$ yields $\eta \sim 0.02$. JPAs on the other hand are operated at $10$\,mK and have been shown to add the minimum amount of noise required by quantum mechanics. In the so-called degenerate mode where only one quadrature is amplified, $\eta$ approaches $1$, whereas $\eta \leq 0.5$ in the phase-preserving mode where both quadratures have equal gain. For the purpose of our proposal, we will assume that $\eta=0.5$, a value which has been obtained in recent experiments~\cite{Campagne.PhysRevX.3.021008(2013)}.

\subsection{Spin-resonator coupling}

We will assume in the following that the spin system we want to measure is subject to a dc magnetic field $\mathbf{B}_0$ applied by an external coil parallel to the resonator inductance in the sample plane (the $z$ direction) in order to minimize its detrimental effect on the superconducting resonator. The spin is described by a Hamiltonian $H_s (\mathbf{B}_0 ) $ that includes a Zeeman term $- \hbar \gamma_e \mathbf{B}_0 \cdot \mathbf{S}$, where $\gamma_e / 2\pi = 28$\,GHz/T is the electron gyromagnetic ratio and $\mathbf{S}$ the dimensionless spin operator, as well as possibly other terms originating either from the hyperfine interaction with one or several nuclear spins, or from some zero-field splitting. In the next sections we will provide two specific examples of such spin Hamiltonians. The resonator is described by its Hamiltonian $H_r/\hbar = \omega_r a^\dagger a$, with field annihilation and creation operators $a$ and $a^\dagger$. Finally, the spin is coupled by the Hamiltonian $H_{int} / \hbar = -  \gamma_e \mathbf{B}_1 \cdot \mathbf{S}$ to the resonator magnetic field $\mathbf{B}_1 = \delta \mathbf{B} (a + a^\dagger)$, where $\delta \mathbf{B}$ denote the microwave field zero-point fluctuations at the spin location.

The spin Hamiltonian $H_s (\mathbf{B}_0 ) $ can be diagonalized, yielding energy states $\ket{n}$ with energies $E_n$. We will assume that $\mathbf{B}_0$ is chosen such that the transition frequency between two of these levels, that we call $\ket{0}$ and $\ket{1}$ in the following, is brought close to the resonator frequency $\omega_r$. Neglecting the other levels, we represent the restriction of the Hamiltonian to the two-level basis $\{\ket{0},\ket{1}\}$ by the Pauli matrices $\sigma_{x,y,z}$. The bare spin Hamiltonian thus writes $H_s/\hbar = - (\omega_s / 2 )\sigma_z$, where $\omega_s = (E_1 - E_0)/\hbar$, and we assume that the phases of $\ket{0}$ and $\ket{1}$ are defined such that the interaction Hamiltonian restricted to the ${\ket{0},\ket{1}}$ basis can be written

\begin{equation}
\label{eq:Hint}
H_{int} / \hbar =g (\sigma_+ a + \sigma_- a^\dagger),
\end{equation}

where

\begin{equation}
\label{eq:g}
g = - \gamma_e \delta \mathbf{B} \cdot \langle 0 | \mathbf{S} | 1 \rangle
\end{equation}

is the spin-resonator coupling constant, $\sigma_+ = | 1 \rangle \langle 0 |$ and $\sigma_- = | 0 \rangle \langle 1 |$. To obtain Eq.~\ref{eq:Hint}, the rotating-wave approximation has been applied to remove non-resonant $\sigma_+ a^\dagger$ and $\sigma_- a$ terms. One sees that the total Hamiltonian $H_s + H_r + H_{int}$ takes a Jaynes-Cummings form, and that cavity Quantum Electrodynamics (CQED) concepts can thus be applied to the spin-resonator system. The coupling constant $g$ is the key parameter of the Jaynes-Cummings model. Efficient detection requires maximizing $g$ while keeping low decoherence rates of both the cavity and the spin. As seen from Eq.~\ref{eq:g}, this requires choosing spin systems and energy levels with large matrix elements, and most importantly designing the resonator for large magnetic field fluctuations $|\delta \mathbf{B}|$.

\subsection{Spin systems}

\begin{figure}[h]
\centering
\includegraphics[width=\hsize]{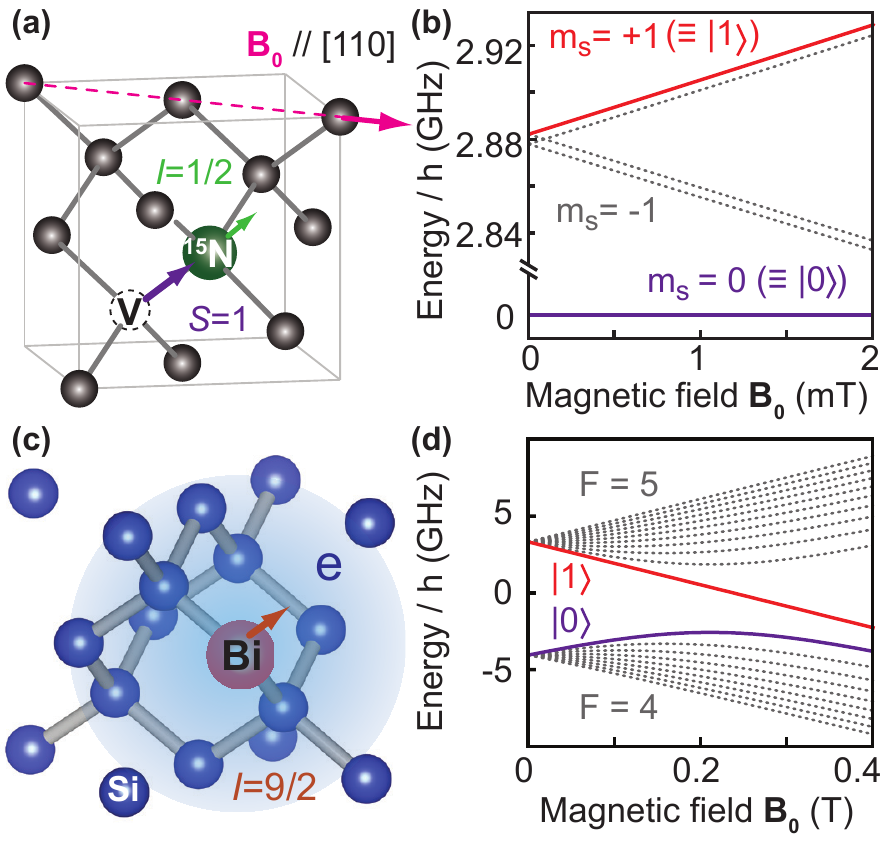}
\caption{The two spin systems studied in this paper.
(a) Schematic of a NV (nitrogen-vacancy) center in diamond crystal.  Here $^{15}$N, which has a nuclear spin $I = 1/2$, is assumed.
(b) Energy levels of $^{15}$NV centers as a function of external magnetic field.  Here the external magnetic field $B_0$ is assumed to be parallel to the orientation of [110], so that a factor $\cos \alpha$ is taken into account (see text).
(c) Schematic of a bismuth donor in silicon (Bi:Si).
(d) Energy levels of Bi:Si as a function of the bias field $B_0$.}
\label{fig:spins}
\end{figure}

In this section, two particular spin systems are considered for implementation of the proposed detection scheme:
nitrogen-vacancy (NV) centers in diamond (Fig.\ref{fig:spins}a.), and bismuth donors in silicon (Fig.\ref{fig:spins}c.).

NV centers are defects in diamond consisting of a nitrogen atom sitting next to a vacancy of the diamond lattice~\cite{Manson.PhysRevB.74.104303(2006)}. In their negatively charged state, the electronic ground state is a spin triplet $S=1$ with a natural quantization axis given by the direction of the $N - V$ bond along one of the four possible $[111]$ directions of the diamond lattice, denoted as $Z$ in the following. We also introduce $X$ and $Y$ as arbitrary axes orthogonal to $Z$. The NV center spin Hamiltonian is then given by

\begin{equation}
\label{eq:HNV}
H_{\mathrm{NV}} (\mathbf{B}_0) / \hbar = D S_{Z}^{2} - \gamma_e \mathbf{B}_0 \cdot \mathbf{S} + A_{Z} I_{Z}S_{Z}.
\end{equation}

The first term is the so-called zero-field splitting ($D / {2 \pi} = 2.88$\,GHz) due to the exchange interaction between the two unpaired electrons of the NV center. The second term is the electronic Zeeman splitting. The last term is the hyperfine interaction with the nitrogen nuclear spin. Here we consider the case of a $^{15}N$ nucleus, which has a spin $I=1/2$ and for which $A_Z/{2\pi} = 3.1$\,MHz. The Hamiltonian Eq.~\ref{eq:HNV} is a good approximation to the full NV Hamiltonian in the limit where the magnetic field $B_0$ obeys $A_Z \ll \gamma_e |B_0| \ll D$, so that both the strain-induced mixing between states $m_S = \pm 1$ and the electron-nuclear spin mixing induced by transverse hyperfine interaction terms have negligible effect; this applies well for magnetic field strengths $B_0$ between $0.1$ and $10$\,mT which we need to tune the spin into  resonance with the resonator frequency $\omega_r$.

Diagonalizing $H_{\mathrm{NV}} (\mathbf{B}_0)$ yields energy eigenstates whose dependence on $B_0$ is shown in Fig.~\ref{fig:spins}. In the magnetic field range that we are interested in, the Hamiltonian can be further approximated by neglecting components of the $\mathbf{B}_0$ field that are transverse to the NV axis, so that $H_{\mathrm{NV}}/ \hbar = D S_{Z}^{2} - \gamma_e B_{0Z} S_Z + A_{Z} I_{Z}S_{Z}$, with $B_{0Z}$ the $\mathbf{B}_0$ component along $Z$. Since this Hamiltonian contains only $S_Z$ and $I_Z$ operators, its eigenstates are of the form $\ket{m_S, m_I}$ and have energies $m_S^2 D - \gamma_e B_{0Z} m_S + A_Z m_S m_I$, $m_S \in \{-1, 0, +1 \}$ and $m_I \in \{-1/2, +1/2 \}$ being the projection of the electronic and nuclear spins along $Z$. Transitions between levels $\ket{m_S, m_I}$ and $\ket{m_S', m_I'}$ verifying $|m_S' - m_S| = 1$ and $m_I' = m_I$ have a non-zero matrix element $\langle m_S' | S_X | m_S \rangle = 1/\sqrt{2}$ and $\langle m_S' | S_Y | m_S \rangle = \pm i /\sqrt{2}$. Note that diamond samples commonly have a surface oriented along the $[100]$ direction. Assuming that the $z$ axis, along which $\mathbf{B}_0$ is applied, is aligned parallel to the $[011]$ crystalline axis, a NV center oriented along the $[111]$ direction experinces a magnetic field $B_{0Z} = ||\mathbf{B}_0|| \cos \alpha$ with $\alpha = 35.3^{\circ}$. For instance, one can define $\ket{0} \equiv \ket{m_S = 0, m_I = +1/2}$ and $\ket{1} \equiv \ket{m_S = -1, m_I = +1/2}$. Assuming $\omega_r /2\pi = 2.9$\,GHz, a field $B_{0Z} = 0.7$\,mT is sufficient to bring the $\ket{0} \rightarrow \ket{1}$ transition into resonance.

%

An important parameter is the spin decoherence rate. Here one needs to distinguish several distinct quantities. First of all, we note that the relaxation rate of an NV center due to exchange of energy with the phonons of the diamond lattice can be neglected at millikelvin temperatures ($\gamma_{\rm dec} < 10^{-3}\,\mathrm{s}^{-1}$~\cite{jarmola_temperature-_2012,grezes_multimode_2014}.)
The only relevant decoherence phenomenon is the loss of phase coherence occurring due to fluctuations of the spin resonance frequency, caused by noise in the magnetic environment experienced by the spin.
It has been extensively studied in the case of individual NV centers~\cite{mizuochi_coherence_2009}.
Two time scales are relevant for our discussion.
The first one is the free-induction decay time $T_2^*$ measured by a Ramsey fringe sequence; it quantifies the time over which the phase of a coherent superposition $(\ket{0} + \ket{1})/\sqrt{2}$ is preserved.
The second one is the time $T_2$ over which a Hahn echo signal decays.
Whereas a Ramsey fringe signal probes very slow fluctuations of the magnetic environment (with a cutoff frequency of the order of $1$\,mHz, determined by the total time of the experiment), a Hahn echo is sensitive to noise with frequencies larger than the inverse of the duration of a single echo sequence, which is of order few kHz.
Because the magnetic environment evolution is usually slow, $T_2 \gg T_2^*$ in general.
NV centers in ultra-pure diamond crystal where most of the carbon atoms are isotopically enriched with nuclear-spin-free $^{12}$C atoms have been shown to reach $T_2^* \gtrsim 400\,\mu \mathrm{s}$ \citep{maurer_room-temperature_2012}.
Hahn-echo decay times in such crystals $T_2$ have been measured up to $5$\,ms at $300$\,K~\cite{Balasubramian.NatMat.8.383(2009)}, where they were limited by the spin-lattice relaxation time, and have been shown to increase at low temperatures~\cite{Bargill.NatCom.4.1743(2013)}.
Note however that these numbers were obtained for NV centers implanted deep in the bulk crystal.
NVs closer to the surface are known to have shorter coherence times, due to the presence of a bath of electron spins of unknown origin at the diamond-air interface \citep{yamamoto_extending_2013}.
Overall, NVs at $15$\,nm from the surface can realistically reach $T_2^* = \gamma_\phi^{-1} = 10 \mu \mathrm{s}$, and $T_2 = 100 \mu \mathrm{s}$ \citep{yamamoto_strongly_2013}.

A neutral bismuth donor in silicon~\cite{Feher1959a,Morley.NatureMat.9.725(2010),wolfowicz_decoherence_2012} (Fig. \ref{fig:spins}c) consists of a single electronic spin $S=1/2$ in hyperfine interaction with the Bi nuclear spin ($I=9/2$), resulting in $20$ hybridized electro-nuclear spin states. The spin Hamiltonian of bismuth in silicon is

\begin{equation}
\label{eq:HBiSi}
H_{\mathrm{Bi}} / \hbar = A \mathbf{I} \cdot \mathbf{S}  - \gamma_e \mathbf{B}_0 \cdot \mathbf{S},
\end{equation}

with $A / 2\pi = 1.48$\,GHz; its eigenstates are shown in Fig.~\ref{fig:spins}. In magnetic fields verifying $B_0 \ll A / \gamma_e$ (i.e. $B_0 \ll 50$\,mT), which we assume is the case here, the dominant term is the hyperfine electron-nuclear interaction. In that limit, the energy levels are well approximated by eigenstates of the total spin operator $\mathbf{F} = \mathbf{I} + \mathbf{S}$, characterized by their $\mathbf{F}^2$ and $F_z$ eigenvalues $F(F+1)$ and $m_F$. They are grouped in two multiplets: $9$ low-energy levels (with $F=4$) and $11$ high-energy levels (with $F=5$). Non-zero $S_x$ and $S_y$ matrix elements are found exclusively between any pair of levels $\ket{F,m_F}$ and $\ket{F',m_F'}$ verifying $|m_F - m_F'| = 1$. Non-zero $S_z$ matrix elements are found exclusively between pairs of levels verifying $m_F = m_F'$. For instance, in $B_0 = 3$\,mT, one gets $\langle F=4, m_F = - 4 | S_x | F=5, m_F = - 5 \rangle = 0.47$, $\langle F=4, m_F = - 4 | S_x | F=5, m_F = - 3 \rangle = 0.07$, $\langle F=4, m_F = - 3 | S_x | F=5, m_F = - 4 \rangle = 0.42$,  and $\langle F=4, m_F = - 4 | S_z | F=5, m_F = - 4 \rangle = 0.3$. The strongest transition, therefore yielding the largest coupling constant to the resonator, is thus the $m_F = - 4 \rightarrow m_F = - 5$; its matrix element is very close to $1/2$, the $S_x$ matrix element of an isolated electron in vacuum. We will thus define $\ket{0} \equiv \ket{F=4, m_F = - 4}$, and $\ket{1} \equiv \ket{F=4, m_F = - 5}$, with a transition frequency that can be tuned to a resonator frequency $\omega_r / 2 \pi = 7.3$\,GHz in a $B_0 = 3$\,mT field.

As for NV centers, energy relaxation of donors in silicon by interaction with the lattice phonons can be entirely neglected at millikelvin temperatures, reaching there also $\gamma_{\rm dec} < 10^{-3}\,\mathrm{s}^{-1}$~\cite{tyryshkin_electron_2011}. The Hahn echo decay time $T_2$ of bismuth donors in isotopically purified silicon has been measured to be between $1$ and $1000$\,ms, depending on the donor concentration and applied magnetic field $B_0$\cite{Weis.APL.100.172104(2012),wolfowicz_atomic_2013,bienfait2015reaching}. The phase coherence time $T_2^*$ of individual bismuth donors in silicon has never been measured so far; however one can rely on results obtained recently with phosphorus donors in a $^{28}$Si substrate, where a remarkably narrow linewidth of $1.8$\,kHz was measured (corresponding to $T_2^* = 300 \mu \mathrm{s}$) for donors located at nanometric distances from the sample surface~\citep{muhonen_storing_2014}. Assuming $\gamma_\phi = 10^4\,\mathrm{s}^{-1}$ therefore seems reasonable.

\begin{table}[h]
\begin{ruledtabular}
\begin{tabular}{| c | c | c | c | c |}

 Spin & $g$(rad/s) & $\kappa(\mathrm{s}^{-1})$ & $\gamma_{\phi}^{-1}$ (s) & $\gamma_p^{-1}$ (s)  \\ \hline
 NV	& $2 \pi \times 6.5 \cdot 10^3$ & $0.9 \cdot 10^5$ & $10^{-5}$ & $3 \cdot 10^{-5}$ \\  \hline
 Si:Bi	& $2 \pi \times 8 \cdot 10^3$ & $2.3 \cdot 10^5$ & $10^{-4}$ & $4.5 \cdot 10^{-5}$  \\
\end{tabular}
\end{ruledtabular}
\caption{Parameters used in this paper to calculate expected signals and measurement time.}
\label{table:para}
\end{table}


\subsection{Resonator design}

The spin-resonator coupling constant $g$ depends on the quantum fluctuations $\delta \mathbf{B}$ of the microwave magnetic field sustained by the resonator mode at the spin location. The quantum fluctuations of the resonator microwave current $\delta i$, which give rise to the magnetic field fluctuations, are linked to the resonator frequency $\omega_r$ and impedance $Z_r = \sqrt{L / C}$ by~\cite{Yurke(1984)}

\begin{equation}
\delta  i = \omega_{r} \sqrt{\frac{\hbar}{2 Z_{r}}}.
\end{equation}

\begin{figure}[h]
\centering
\includegraphics[width=\hsize]{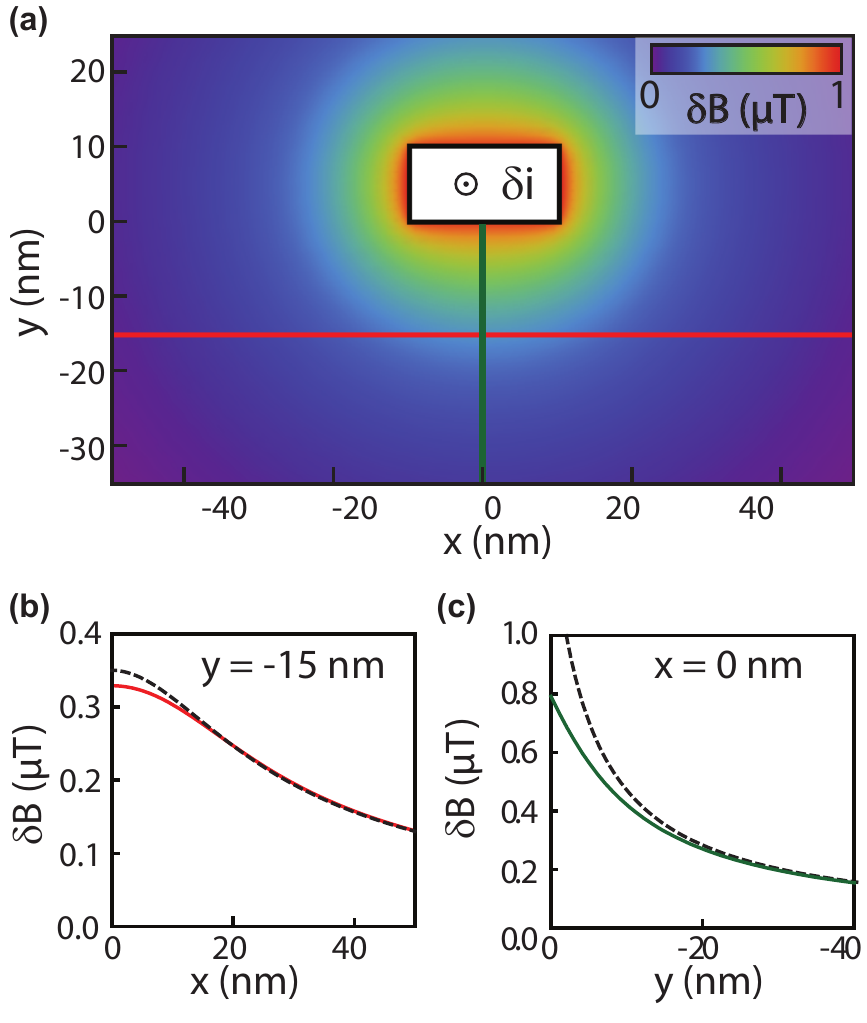}
\caption{Magnetic field $\delta B$ generated by vacuum fluctuations of the current $\delta i$ in a constriction of width 20\,nm and thickness 10\,nm.
(a) Magnetic field map analytically derived from Biot and Savarts law for a conductor with uniform current density and rectangular cross-section. $\delta B$ is given for $\delta i=35\,$nA, as in the NV center case.
(b) Cut at distance 15 nm from the constriction. For $\delta i=35\,$nA, we can expect $\delta B = 0.33 \,\mu$T. Red line is the result of the exact analytical formula, black dashed line shows the approximation $\mu_0  \delta i / (2 \pi \sqrt{x^2 + y^2})$ (c) Cut along vertical crossing through the constriction. Green solid line is the exact formula, black dashed line is $\mu_0  \delta i / (2 \pi \sqrt{x^2 + y^2})$.}
\label{fig:magnField}
\end{figure}

A simple estimate of $|\delta \mathbf{B}|$ is obtained by assuming a circular cross section of the nanowire, in which case Biot and Savart's law yields an orthoradial $\delta \mathbf{B}$ with the amplitude $\delta B = \mu_0  \delta i / (2 \pi r) $ at the spin location at a distance $r$ from the center of the nanowire. Analytical results exist also for a wire of rectangular cross-section as shown in Fig.~\ref{fig:magnField} while in general geometries, the field can be computed with finite element methods. Since the resonator frequency $\omega_r$ should be chosen close to the spin transition frequency, maximizing $\delta B$ requires reducing as much as possible the resonator impedance $Z_r$ and bringing the spin as close as possible to the resonator, i.e. minimizing $r$.

For practical and physical reasons the wire cannot be designed with a width much below $\simeq 20$\,nm as electron-beam lithography can only yield reproducible nanowires with a width larger than $\simeq 15 - 20$\,nm; and nanowires with transverse dimensions below $\simeq 10$\,nm may undergo a superconducting-to-insulating transition~\cite{Zgirski.NanoLett.5.1029(2005)} which would be detrimental to the resonator quality factor. Bringing the spin too close to the surface may lead to reduced coherence times, as has been demonstrated with NV centers in diamond at a depth less than $\simeq 20$\,nm from the surface. Taking these aspects into account, a nanowire width of $20$\,nm, thickness of $15$\,nm, and a spin - nanowire distance $r = 20$\,nm seem appropriate parameters, as proposed in~\cite{Tosi.AIPAdv.4.087122(2014)}.

\begin{figure}[h]
\centering
\includegraphics[width=\hsize]{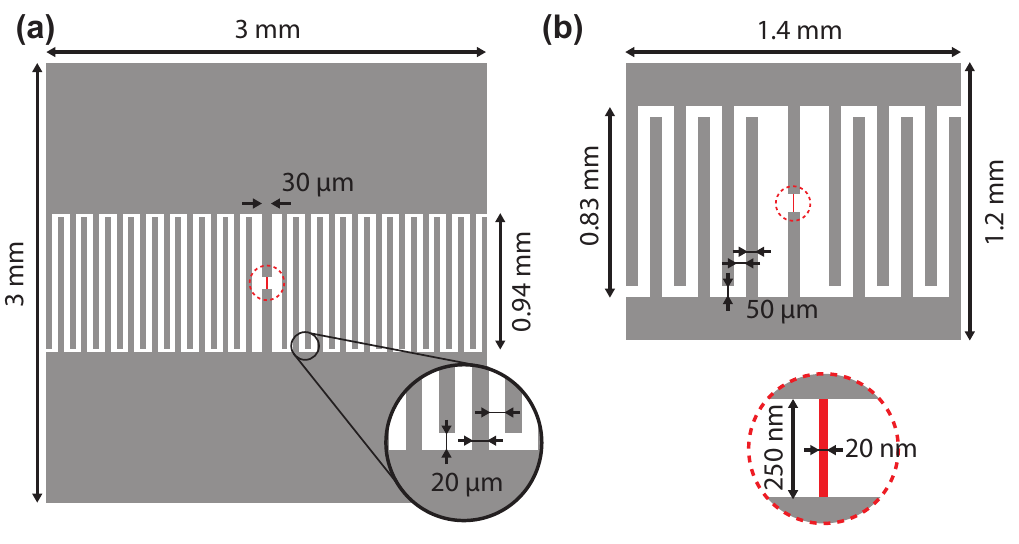}
\caption{Resonator geometries, with the nanowire depicted in red.
(a) for an NV center spin: central wire is $30\,\upmu$m wide and $0.96\,$mm long. Two large pads of $3$mm by $1$mm ensure the coupling to the cavity and $72$ pairs (only half are drawn) of $20\,\upmu$m-wide fingers spaced by $20\,\upmu$m are used as additional capacitance to bring the impedance down to $Z_r=15.3\Omega$. (b) for bismuth in silicon. Central wire is $50\,\upmu$m wide and $0.83\,$mm long. Two large pads of $1.4$mm by $0.18$mm ensure the coupling to the cavity and $6$ pairs of $50\,\upmu$m-wide fingers spaced by $50\,\upmu$m are used as additional capacitance to bring the impedance down to $Z_r=26.5\Omega$.}
\label{fig:resos}
\end{figure}

The resonator design aims at minimizing its impedance, which implies maximizing the capacitance $C$ while minimizing the inductance $L$. We propose to use an interdigitated capacitor, which is known to be compatible with high quality factor resonances required by the experiment provided the finger dimension and spacing is large enough (above $10-20 \mu \mathrm{m}$~\cite{Megrant.APL.100.113510(2012)}), in parallel with an inductor made out of a straight superconducting wire. Large pads facilitate the capacitive coupling to the antenna. To minimize the impedance, the width of the inductive wire should be as large as possible outside of the nanowire region. The total resonator inductance should also include the kinetic inductance of the nanowire, which can be evaluated as~\cite{Lk_formula}:

$$ L_k= \frac{l}{w} \frac{R_\square}{2\pi^2} \frac{h}{\Delta}\frac{1}{\tanh{\frac{\Delta}{2 k_B
T}}} $$

where $\Delta$ is the aluminum superconducting energy gap, $T$ the temperature, $R_\square$ the sheet resistance in the non-superconducting state, and $l,w$ the nanowire length and width. For a 10-nm-thick aluminum film, $R_\square=4.5 \Omega$ \cite{ThinFilm_Resistivity_DeVries1988} and $\Delta=230 \upmu$eV \cite{AlThinFilmGap} yield $L_k=50\,$pH for $l=250$\,nm and $w=20$\,nm.

The proposed resonator geometry for the NV centers is shown in Fig.\ref{fig:resos}(a). A geometrical inductance $L_g = 790$\,pH is achieved by using a $30$-$\upmu$m-wide central wire, while the capacitor includes $36$ pairs of $20 \mu \mathrm{m}$-wide fingers separated by $20 \mu \mathrm{m}$. With the nanowire modelled as an ideal inductor $L_k = 50$\,pH, electromagnetic simulations give $\omega_r / 2 \pi = 2.9$\,GHz and $Z_r = 15.3\,\Omega$, yielding $\delta i = 35$\,nA. For a distance of $15$\,nm, we get a field $\delta B = 0.33 \,\mu$T (see Fig.~\ref{fig:magnField}). Using the $\ket{0} \rightarrow \ket{1}$ transition described earlier, one obtains a coupling constant $g / 2\pi = 6.5$\,kHz if the NV center is positioned right below the wire so that $\delta \mathbf{B}$ is perpendicular to the NV axis $Z$.

A similar geometry is proposed for coupling to individual bismuth donors in silicon (see Fig.~\ref{fig:resos}(b)). There the geometrical inductance $L_g = 530$\,pH is obtained with a $50 \mu \mathrm{m}$-wide wire, and the capacitor includes $6$ pairs of $50 \mu \mathrm{m}$-wide fingers separated by $50 \mu \mathrm{m}$. This yields $\omega_r / 2\pi = 7.3$\,GHz and $Z_r = 26.5\,\Omega$, implying $\delta i = 65$\,nA, resulting in  a field $\delta B = 0.61 \,\mu$T at the spin location. With the choice of levels $\ket{0}$ and $\ket{1}$ described in the previous paragraph, we get $g / 2\pi = 8.0$\,kHz.

%
\section{The master equation and steady state signal amplitude}
\label{sec:framework}

In this section we shall determine the steady state of the spin-resonator system, and we shall quantify the dependence of the amplitude of the field emitted by the resonator on the physical parameters of the problem. To this end, we must establish the master equation, which is of the general Lindblad form,
\eq \label{eq:ME}
d\rho=-\frac{i}{\hbar}[H, \rho]dt+\sum_j\mathcal{D}[c_j]\rho dt.
\eeq
The Hamiltonian of the spin, interacting with a coherently driven resonator, can be written in a frame rotating with the driving field, $\beta_\text{in}=\beta e^{-i \omega_d t}$,
\eqa
H &=& \hbar\Delta_ ra^\dg a_r + i\hbar\sqrt{2\kappa_1}(\beta a^\dg-\beta^*a)+\frac{\hbar\Delta_s}{2}\sz \nn\\
&&+ \hbar g (\spl a + \sm a^\dg),
\eeqa

where $\Delta_{r(s)} = \omega_{r(s)}-\omega_d$ is the detuning between the resonator (spin) and the driving frequency. The field inside the resonator is described by creation and annihilation operators $a$ and $a^\dg$, and $\kappa_1$ is the damping rate of the resonator field through the coupler. The total damping rate of the resonator field $\kappa = \kappa_1 + \kappa_L$ taking into account internal resonator losses with rate $\kappa_L$ is linked with the total resonator quality factor $Q$ by $\kappa = \omega_r / (2 Q)$. We obtain information about the presence of the spin by detection of the field leaking with damping rate $\kappa_1$ from the resonator through the coupler.

All dissipation processes are treated in the Born-Markov approximation in (\ref{eq:ME}) with Lindblad master equation terms of the form
\eq
\mathcal{D}[c]\rho=  c\rho c^\dg - \frac{1}{2}\left\{c^\dg c,\rho\right\}.
\eeq
The relevant damping processes are the decay of the resonator field, $c_1 = \sqrt{2\kappa}a$, population decay of the spin, $c_2 = \sqrt{\gamma_\text{dec}}\sm$, and spin dephasing, $c_3 = \sqrt{\frac{\gamma_\phi}{2}}\sz$.

\subsection{Adiabatic elimination and steady state expectation values}

Due to the relatively weak value of the spin-resonator coupling constants derived in paragraph~\ref{sec:model}, one can assume that the resonator lifetime is the shortest time scale of the problem ($\kappa\gg g$), and that the resonator field closely follows the values of the driving field and the spin coherence. In this so-called bad-cavity limit, one can adiabatically eliminate the resonator field mode to obtain an effective master equation for the spin degrees of freedom \cite{julsgaard_measurement-induced_2012}.

The spin-resonator Hamiltonian yields the Heisenberg equation of motion for the resonator annihilation operator,
\begin{equation} \label{eq:Heis-a}
\dot{a} = -i\Delta_r a + \sqrt{2\kappa_1}\beta - ig\sigma_- - \kappa a + \hat{F},
\end{equation}
where the damping term, $-\kappa a$ follows by incorporating a non-Hermitian term $-i\hbar\kappa a^\dg a$ in the Hamiltonian, and where $\hat{F}$ is a Langevin noise term with vanishing expectation value.

In the absence of the spin, the resonator will be excited into a coherent state with a steady state amplitude $\alpha$ that follows by taking expectation values on both sides of~(\ref{eq:Heis-a}) and setting the time derivative to zero,
\begin{equation}
\alpha = \frac{\sqrt{2\kappa_1}\beta}{\kappa + i \Delta_r}.
\end{equation}

The spin perturbs the field only weakly and, following \cite{julsgaard_measurement-induced_2012}, we shall write the resonator field operator as $a = \alpha + a'$, where the Heisenberg equation of motion for the operator $a'$ follows  from Eq.~(\ref{eq:Heis-a}). Since the spin operator term in the equation evolves at the natural frequency $\omega_s$, we assume the adiabatic following (vanishing time derivative of $a'$) in a frame rotating at that frequency. This yields the operator replacement, $a' = \frac{-ig\sigma_-}{\kappa + i\Delta_{rs}}$, where $\Delta_{rs}=\Delta_r - \Delta_s$, and hence the expression for the total resonator field operator,
\begin{equation} \label{eq:ad-val}
a = \frac{\sqrt{2\kappa_1}\beta}{\kappa + i \Delta_r} - \frac{ig\sigma_-}{\kappa + i\Delta_{rs}}.
\end{equation}

Inserting this expression and its adjoint for $a^\dg$ in the original Hamiltonian and Lindblad operators, we obtain an effective master equation involving only the spin degree of freedom.

The system is thus governed by an effective spin Hamiltonian,
\eq
H_\text{eff} = \frac{\hbar\Delta_s}{2}\sz+\hbar g (\alpha\spl + \alpha^*\sm)-\hbar\epsilon_s\spl\sm\label{Heff},\nn\\
\eeq
where $\epsilon_s = \Delta_{rs}g^2/(\kappa^2+\Delta_{rs}^2)$ denotes a small AC-Zeeman-like shift of the spin energy levels.

Similarly, the damping terms become
\eq
c_{1,\text{eff}}=\sqrt{\gamma_p}\sm,\, c_2  =  \sqrt{\gamma_\text{dec}}\sm,\, c_3  = \sqrt{\frac{\gamma_\phi}{2}}\sz,\nn
\eeq
where the rate $\gamma_p=2g^2\kappa/(\kappa^2+\Delta_{rs}^2)$ represents the Purcell enhanced damping of the spin by spontaneous emission of a photon into the output line due to the coupling to the cavity mode. This cavity-enhanced decay is an essential point of our proposal: due to the coupling to the cavity mode, the signal reflected by the resonator will be appreciably influenced by the spin, and as we shall see below, the Purcell rate $\gamma_p$ is the crucial parameter for the sensitivity of the scheme. We note that cavity-enhanced spin relaxation was observed recently~\cite{bienfait2015controlling}.

From the simple two-level master equation of the spin we find the steady state value
\eqa \label{eq:sigmam}
\langle \sm\rangle_\text{ss}&=&-\frac{ig\alpha_r\gamma_1\gamma_2^*}{4g^2|\alpha|^2\text{Re}(\gamma_2)+\gamma_1|\gamma_2|^2},
\eeqa
with $\gamma_1 = \gamma_{\text{dec}}+\gamma_p$ and $\gamma_2 = \gamma_1/2+ \gamma_\phi-i(\Delta_s-\epsilon_s)$.
As explained in section II, in the situations that we consider here the spin population decay rate $\gamma_{dec}$ is negligible compared to the Purcell rate $\gamma_p$~\cite{bienfait2015controlling} and we shall hence assume $\gamma_1 = \gamma_p$.

The largest modification of the intra-cavity field (\ref{eq:ad-val}) due to the spin is found when the resonator is on resonance with both the spin and the driving field, $\Delta_r=\Delta_s=\epsilon_s =0$, with the corresponding decay rates $\gamma_p = 2g^2/\kappa$ and $\gamma_2=\gamma_1/2+\gamma_\phi$, and when $|\alpha| = |\alpha|_\text{sat}=\sqrt{\gamma_1\gamma_2}/2g = \sqrt{\gamma_2/\kappa}$. In the rest of the article we will assume this optimal regime for which we find the expression,

\eqa
\langle \sm\rangle_\text{ss}&=&- \frac{i \sqrt{2}}{4} \frac{g}{\sqrt{\gamma_2 \kappa}}. 
\eeqa
%
The modification to the steady state cavity field (assuming $\kappa_L \ll \kappa_1$) is then given by

\eqa
\sqrt{2 \kappa} \langle a'\rangle_\text{ss}&=&- \frac{ 1}{2} \frac{g^2}{\sqrt{\gamma_2} \kappa}. 
\eeqa

While the spin energy decay rate is negligible compared to the Purcell rate $\gamma_p$, this is not necessarily the case for the spin dephasing rate $\gamma_\phi$, which brings us to distinguish two limiting cases. If the spin coherence is radiatively limited ($\gamma_p \gg \gamma_\phi$, implying that $\gamma_2 = \gamma_p/2$), one obtains $\langle \sm\rangle_\text{ss}=- \frac{i \sqrt{2}}{4}$ and

\begin{equation}
\sqrt{2 \kappa} \langle a'\rangle_\text{ss} = - \frac{g}{2 \sqrt{\kappa}} 
\end{equation}

If instead the spin coherence time is limited by dephasing so that $\gamma_2 = \gamma_\phi$, we get 

\begin{equation}
\sqrt{2 \kappa} \langle a'\rangle_\text{ss} = - \frac{g^2}{2 \sqrt{\gamma_\phi}\kappa}.
\end{equation}

\subsection{Detection of the microwave field reflected by the resonator}

The microwave signal reflected by the resonator has the operator expression,

\begin{eqnarray} \label{eq:ref}
c_{out}&=& \sqrt{2\kappa_1}(\alpha - \frac{ig}{\kappa+i\Delta_{rs}}\sigma_-) - \beta \nonumber \\
&=& (\frac{2\kappa_1}{\kappa+i\Delta_r} -1)\beta - \frac{i\sqrt{2\kappa_1} g}{\kappa+i\Delta_{rs}}\sigma_-.
\end{eqnarray}

This signal is amplified and demodulated to yield a voltage signal, similar to the one obtained in optical homodyne detection,
\eq \label{eq:dY}
dY = \eta\, \langle c_m+c_m^\dg\rangle dt+ \sqrt{\eta}\,dW,
\eeq
composed of a mean value governed by the expectation value of the output field,
\eq \label{eq:cm}
c_{m} =c_{out}  e^{-i\theta},
\eeq
where the local oscillator phase $\theta$ is applied to choose the appropriate quadrature component measured by the set-up. The Wiener noise term  $dW$ has zero mean and variance $dt$, and it represents detector shot noise. The parameter $\eta \leq 1$ denotes the detector efficiency introduced in paragraph~\ref{sec:model}.

\begin{figure}
\begin{center}
\includegraphics[width=0.47\textwidth]{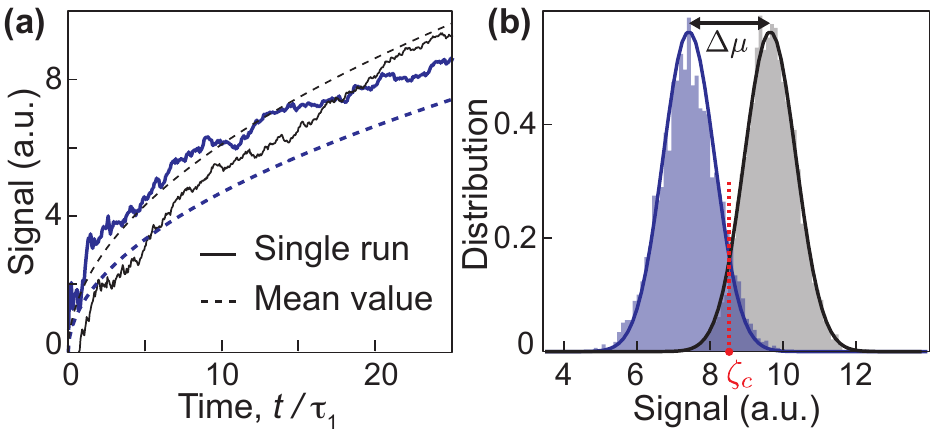}
\caption{Integrated currents when there is a spin interacting with the resonator (denoted with blue) and when there is no spin (denoted with grey). The simulated voltage signals were obtained with the parameters: $g=2\pi\times 10 \cdot 10^3$ (rad/s), $\kappa=4.6 \cdot 10^5$ (s$^{-1}$), $\gamma_{\phi} = 10^{4}$ (s$^{-1}$), and $\eta=0.5$. In panel (a) the solid lines correspond to values of the integrated current for single simulations and the dotted lines to mean values for 3000 simulations. The distributions of the ensembles of simulations at time $t=20\ \tau_1 \simeq 5$ ms are shown in panel (b), where the solid lines are Gaussian fits to each distribution. The threshold value $\zeta_c$ is denoted with the vertical dotted line.}
\label{fig:Gaussians}
\end{center}
\end{figure}

We choose the phase of the driving field such that $\alpha$ is real and assume that all but a negligible fraction of the photons lost from the resonator are available for homodyne detection, $\kappa\approx\kappa_1$. In that case (and assuming vanishing detuning parameters), $c_{out} = \beta - i\sqrt{\gamma_p}\sigma_-$, and it is convenient to introduce the normalized integrated signal
\eqa \label{eq:current}
\zeta(t) &=& \frac{1}{\sqrt{t}}\int_0^t dY \nn\\
&\approx& 2\eta\cos\theta\sqrt{t}\left(\beta-i\sqrt{\gamma_p}\langle \sm\rangle_\text{ss}\right) + \sqrt{\eta}\Delta W,
\eeqa
where $\Delta W$ is a Gaussian distributed noise term with zero mean and unit variance. The mean integrated signal is maximal when $\theta = 0$. For short times, it is dominated by the noise fluctuations while for longer times, the integrated currents differ by more than the fluctuations and enable discrimination of whether the spin is present in the resonator or not.

In Fig. \ref{fig:Gaussians} we show simulations of the noisy integrated currents corresponding to measurement signals from a resonator interacting with a spin, and to measurement currents with no spin. Typical trajectories obtained by the stochastic master equation, presented in Sec. IV, are shown for both scenarios as the solid blue and black lines, respectively. The distributions of the integrated signals (\ref{eq:current}) can be used to estimate the error in assigning a given value of the integrated signal to the spin or the no spin hypothesis. Fig. \ref{fig:Gaussians} (b) shows that the integrated currents, normalized by $\sqrt{t}$, are Gaussian distributed with a constant variance $\eta$. The mean values evolve as $\sim\sqrt{t}$ (dotted lines in Fig. \ref{fig:Gaussians} (a)) and the separation of the two distributions is given by
\eq
\Delta\mu = 2\eta\sqrt{\gamma_p t} |\langle\sm\rangle_\text{ss}|.
\eeq
For the parameters leading to Eq.(13), we thus obtain a unit signal-to-noise ratio by integrating the signal for a duration of
$\tau_{\eta} = \frac{1}{\eta} \tau_{1}$, where

\begin{equation} \label{eq:tau1}
\tau_{1} \equiv \frac{1}{4\gamma_p |\langle\sm\rangle_\text{ss}|^2}  = \frac{\kappa^2 \gamma_2}{g^4}.
\end{equation}

The Gaussian fits to the histograms in Fig.5(b) are in perfect agreement with our theoretical analysis. The variances of the curves are equal to $\eta=0.5$ as predicted by Eq.(20). For $t=20 \tau_1 = 10 \tau_\eta$, we expect a factor of $\sqrt{10}$ between the separation and the r.m.s. width of the Gaussians, i.e., a separation of $\sqrt{10\eta}=\sqrt{5}\simeq 2.24$, which is  also the result of Eq.(21), and which  perfectly fits the simulations. Given the integrated signal from an experiment, we conclude that we are (not) coupled to a spin, if the signal is larger (smaller) than a threshold value $\zeta_c$, the mid-point of the two peaks in \ref{fig:Gaussians} (b). The probability that this assignment is in error is given by the area under the Gaussian tails beyond $\zeta_c$, which we can evaluate as function of the probing duration for any value of the detector efficiency,
\eq \label{eq:error1}
\epsilon_{\eta}(t) = \frac{1}{2}\left[1-\text{erf}\left(\frac{\sqrt{\eta}}{2\sqrt{2}}\sqrt{\frac{t}{\tau_1}}\right)\right],
\eeq
where $\text{erf}$ is the Gaussian error function. This error vanishes exponentially in the limit of long measurement times, and it is shown by the smooth curves in Fig.~\ref{fig:errors} for different values of the measurement efficiency. The noisy curves in the same figure depict the error probability associated with a Bayesian trajectory analysis of the full measurement record, which will be discussed in the following section.

At this point, it is interesting to evaluate the parameters for the cases of NV centers in diamond and bismuth donors in silicon. With the figures provided in section II, NV centers have a Purcell relaxation time $\gamma_p^{-1} = 27 \mu \mathrm{s}$, yielding a measurement time $\tau_{1} = 0.35 $\,ms (assuming unit detector efficiency $\eta=1$). Similarly, bismuth donors in silicon can reach $\gamma_p^{-1} = 45 \mu \mathrm{s}$, resulting in $\tau_{1} = 0.17$\,ms. We thus conclude that a high fidelity single spin microwave detection should be possible in just milliseconds for the model systems considered in this work.

\section{Quantum trajectories and Bayesian analysis}
\label{sec:bayes}

In the previous section we showed that the value of the voltage signal integrated over a few $\frac{1}{\eta}\tau_1$ allows discrimination of the spin in the resonator.
By representing the signal by only its integral over time, however, we omit important information contained in the temporal signal correlations. We can assess this information by application of quantum trajectory theory, which evolves the quantum state in time, conditioned on the stochastic measurement record. This state in turn provides the probabilities for subsequent values of the detected signal. The outcome probabilities depend on whether the spin is included in the simulation or not, and given the actual outcome, we can apply Bayes' rule and infer the (classical) probability that the spin is actually present.

The state of a quantum system subject to continuous probing obeys a stochastic master equation (SME)
\eq \label{eq:SME}
d\rho=-\frac{i}{\hbar}[H, \rho]dt+\sum_j\mathcal{D}[c_j]\rho dt+ \sqrt{\eta}\mathcal{H}[c_m]\rho\,dW,
\eeq
where the stochastic term, which has been added to the conventional master equation (\ref{eq:ME}) accounts for the back action of the noisy measurement with the outcome $dY$ given in  (\ref{eq:dY}). Note that $dW$ is determined as the difference between the actually measured signal $dY$ and the expected mean value from the current value of the density matric $\rho$.  The back action involves  application of the
superoperator
\eq \label{eq:measurement}
\mathcal{H}[c_m]\rho = c_m\rho + \rho c_m^\dg - \langle c_m+c_m^\dg\rangle\rho
\eeq
with $\langle c_m+c_m^\dg \rangle = \text{tr}\left\{(c_m+c_m^\dg)\rho\right\}$, and with the operator $c_m$ given in Eq.(\ref{eq:cm}).

The SME thus yields the time dependent state $\rho(t)$ of the system conditioned on the measurements until time $t$, and it provides the expected mean value and the probability for the next detector outcome. Propagating the SME for different candidate hypotheses or parameter values thus provides the necessary input to apply Bayes rule and determine the most likely circumstance of the experiment:
The actual measurement outcome $dY$ during a time interval $[t,t+dt]$ updates the probability $p(\theta_i)$ that an unknown parameter has a certain value $\theta_i$ via Bayes' rule:
\eq \label{eq:bayes}
p(\theta_i, t+dt)\equiv p(\theta_i|dY) = \frac{p(dY|\theta_i)}{p(dY)} p(\theta_i,t),
\eeq
where $p(\theta_i,t)$ is the \emph{prior} probability of the parameter having value $\theta_i$ at time $t$. The denominator $p(dY) = \sum_ip(dY|\theta_i)p(\theta_i,t)$ merely serves to normalise the updated probabilities so that $\sum_i p(\theta_i|dY)=1$. After each infinitesimal time-step $dt$ we thus update the probability distribution $\{p(\theta_i, t)\} \rightarrow \{p(\theta_i, t+dt)\}$, which evolves during the full measurement process. If we start, for example, with a uniform distribution $p(\theta_i, 0) = 1/N$, where $N$ is the number of candidate hypotheses, we must, in parallel, solve $N$ stochastic master equations, which are all subject to the same measurement record $dY$. Each solution thus provides $p(dY|\theta_i)$, needed in (\ref{eq:bayes}) to update the probability weights on the different hypotheses.

\begin{figure}
\begin{center}
\includegraphics[width=0.47\textwidth]{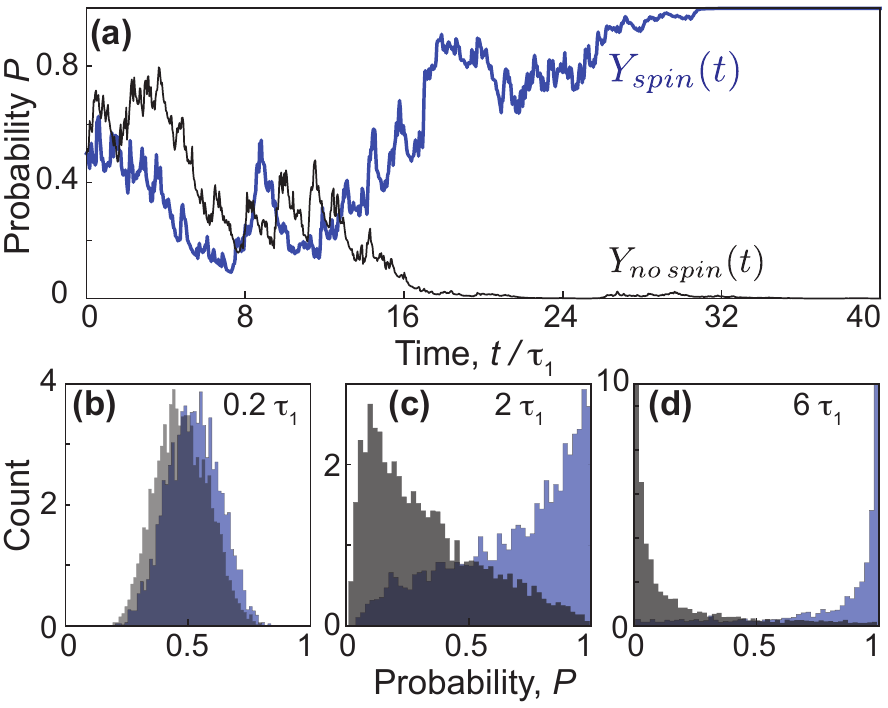}
\caption{The probability of inferring that a spin is interacting with the superconducting resonator, $P \big(\text{spin}|Y(t) \big)$. Panel (a) shows the probabilities as a function of time when $dY(t)$ is a measurement current from a simulation with a spin (blue curve) and when there is no spin (black curve). Panels (b) - (d) show the normalised distribution of the values of $P \big(\text{spin}|Y(t) \big)$ for 3000 simulations at measurement times $t=0.2, 2$ and $6$ $\tau_1$, respectively. Blue histograms correspond to simulations with a spin and black histograms to simulations with no spin interacting with the resonator. The parameters of the simulation are specified in Fig.(\ref{fig:bayes}).}
\label{fig:bayes}
\end{center}
\end{figure}

To assess the efficacy of the Bayesian analysis for our purpose, we have assigned equal \emph{prior} probabilities for having a spin and having no spin in the resonator, $p(\text{spin}) = p(\text{no spin}) = 1/2$, and simulated measurement currents corresponding to the two cases. The simulated voltage signals were obtained with the parameters: $g=2\pi\times 10 \cdot 10^3$\,rad/s, $\kappa=4.6 \cdot 10^5 \mathrm{s}^{-1}$, $\gamma_{\phi} = 10^{4} \mathrm{s}^{-1}$, and $\eta=0.5$. We used Bayes' formula  (\ref{eq:bayes}) to update the probabilities that a person having only access to the measurement data would assign to the two possibillies of having a spin or no spin in the resonator. The probabilities evolve with time as shown with the blue and  black curve in Fig. \ref{fig:bayes} (a) for two distinct simulations with signals generated as if a spin is present or not. The black curve rapidly converges to value zero, deducing correctly that the resonator is not interacting with a spin. The blue curve fluctuates for slightly longer, but eventually reaches unit probability around a measurement time $t=18-20 \tau_1 \simeq 5$\,ms with our parameters.

\begin{figure}
\begin{center}
\includegraphics[width=0.4\textwidth]{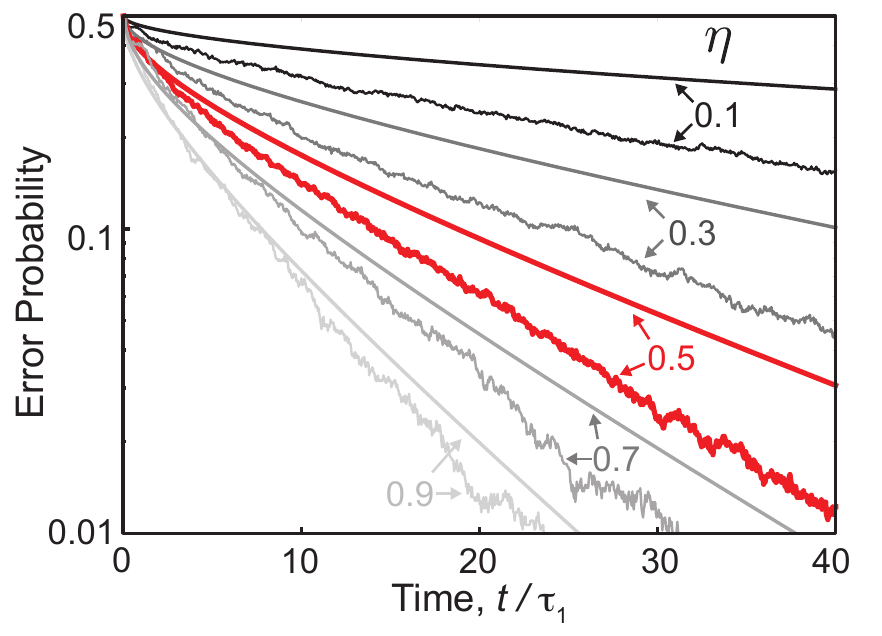}
\caption{The time dependent probability of wrongly assigning the presence or absence of a spin based on the continuous measurement. The probabilities are shown on a logarithmic scale for different values of the detector efficiency $\eta$ indicated next to the arrows which connect the smooth curves depicting Eq. (\ref{eq:error1}), based on the integrated measurement current, and the more noisy curves, based on the Bayesian analysis. The red smooth and noisy curves correspond to the experimentally realistic detector efficiency value $\eta=0.5$. The interaction and damping parameters are the same as in Fig.(\ref{fig:bayes}).}
\label{fig:errors}
\end{center}
\end{figure}

We have repeated the simulations 3000 times to obtain the distribution of probabilities, shown for  measurement times $t=0.4, 4$ and $12$ $\tau_1$ in Fig. \ref{fig:bayes} (b)-(d). The blue histograms correspond to simulated measurement records with a spin interacting with the resonator and the black ones to the case of no spin. As in our analysis based on the integrated signals, we can assess the probability of making a wrong assignment by the tail of the blue (black) distributions extending above (below) the conditional probability  $p=0.5$. Unlike in the previous section, this assignment error probability does not have an analytical expression, but it can be determined from our numerical simulations. For different detector efficiencies we thus obtain the assignment errors shown as the noisy curves in Fig.~\ref{fig:errors}.

As shown in Fig.~\ref{fig:errors}, the Bayesian analysis extracts more information from the measurement signal than we can obtain based on the integrated signal. The difference in the error probability is quite appreciable and for example, allows a Bayesian analysis of data obtained by a detector with efficiency $\eta=0.5$ to yield the same information as the integrated signal obtained with a detector efficiency of $0.65$. Alternatively, we observe that with the same detector efficiency we obtain a confident discrimination of the presence of a spin 30\,\% faster by use of the Bayes analysis than by use of only the integrated signal.

The improvement due to the Bayesian analysis is due to the information retrievable from temporal correlations in the emitted signal. The random measurements cause back action on the system and thus cause its subsequent transient evolution to deviate at all times from the constant steady state of the master equation. Such transient evolution yields more information than mean values about the system as seen most dramatically, e.g., in photon counting signals, where a two-level emitter makes a quantum jump into the ground state after each detection event, and hence has to be re-excited before a second detection can occur. While the mean, and hence integrated, counting signal saturate and thus hamper distinction between different strong driving fields, the transient excited state population after each detector click causes anti-bunching and a modulation in the intensity correlation function which, in fact, allows the same resolution at all driving strengths~\cite{Kiilerich.PhysRevA.89.052110(2014)}.

For continuous homodyne detection, the stochastic back action on the emitter is a weaker effect than in the case of counting, but it still accounts for temporal correlations in the noisy signal which are not being used in the mean field analysis, e.g., the frequency spectrum of the emitted field~\cite{Kiilerich.PhysRevA.91.012119(2015)}.

\section{Conclusion \label{sec:conclusions}}

In summary, we have proposed and analyzed an experimental scheme where a superconducting microwave resonator is employed for detection of a single electron spin. From the steady state of the system we identified the dependence of the mean signal and the signal fluctuations on the spin coupling parameters and we have shown that with realistic parameters, we can detect the presence of a single spin within an integration time of few milliseconds. We also showed that a Bayesian analysis of the detected signal permits faster and more reliable discrimination of the spin than the mean field analysis due to the temporal signal correlations associated with the measurement back action on the spin dynamics.

\begin{acknowledgments}
The authors acknowledge discussions with Alexander Holm Kiilerich, Brian Julsgaard, John Morton, Jarryd Pla, and within the Quatronics group. We acknowledge funding from the Villum Foundation and from the European Research Council under the European Community's Seventh Framework Programme (FP7/2007-2013) through grant agreement No. 615767 (CIRQUSS).
\end{acknowledgments}


\begin{thebibliography}{59}%
\makeatletter
\providecommand \@ifxundefined [1]{%
 \@ifx{#1\undefined}
}%
\providecommand \@ifnum [1]{%
 \ifnum #1\expandafter \@firstoftwo
 \else \expandafter \@secondoftwo
 \fi
}%
\providecommand \@ifx [1]{%
 \ifx #1\expandafter \@firstoftwo
 \else \expandafter \@secondoftwo
 \fi
}%
\providecommand \natexlab [1]{#1}%
\providecommand \enquote  [1]{``#1''}%
\providecommand \bibnamefont  [1]{#1}%
\providecommand \bibfnamefont [1]{#1}%
\providecommand \citenamefont [1]{#1}%
\providecommand \href@noop [0]{\@secondoftwo}%
\providecommand \href [0]{\begingroup \@sanitize@url \@href}%
\providecommand \@href[1]{\@@startlink{#1}\@@href}%
\providecommand \@@href[1]{\endgroup#1\@@endlink}%
\providecommand \@sanitize@url [0]{\catcode `\\12\catcode `\$12\catcode
  `\&12\catcode `\#12\catcode `\^12\catcode `\_12\catcode `\%12\relax}%
\providecommand \@@startlink[1]{}%
\providecommand \@@endlink[0]{}%
\providecommand \url  [0]{\begingroup\@sanitize@url \@url }%
\providecommand \@url [1]{\endgroup\@href {#1}{\urlprefix }}%
\providecommand \urlprefix  [0]{URL }%
\providecommand \Eprint [0]{\href }%
\providecommand \doibase [0]{http://dx.doi.org/}%
\providecommand \selectlanguage [0]{\@gobble}%
\providecommand \bibinfo  [0]{\@secondoftwo}%
\providecommand \bibfield  [0]{\@secondoftwo}%
\providecommand \translation [1]{[#1]}%
\providecommand \BibitemOpen [0]{}%
\providecommand \bibitemStop [0]{}%
\providecommand \bibitemNoStop [0]{.\EOS\space}%
\providecommand \EOS [0]{\spacefactor3000\relax}%
\providecommand \BibitemShut  [1]{\csname bibitem#1\endcsname}%
\let\auto@bib@innerbib\@empty
\bibitem [{\citenamefont {Bar-Gill}\ \emph {et~al.}(2013)\citenamefont
  {Bar-Gill}, \citenamefont {Pham}, \citenamefont {Jarmola}, \citenamefont
  {Budker},\ and\ \citenamefont {Walsworth}}]{Bargill.NatCom.4.1743(2013)}%
  \BibitemOpen
  \bibfield  {author} {\bibinfo {author} {\bibfnamefont {N.}~\bibnamefont
  {Bar-Gill}}, \bibinfo {author} {\bibfnamefont {L.}~\bibnamefont {Pham}},
  \bibinfo {author} {\bibfnamefont {A.}~\bibnamefont {Jarmola}}, \bibinfo
  {author} {\bibfnamefont {D.}~\bibnamefont {Budker}}, \ and\ \bibinfo {author}
  {\bibfnamefont {R.}~\bibnamefont {Walsworth}},\ }\href {\doibase
  10.1038/ncomms2771} {\bibfield  {journal} {\bibinfo  {journal} {Nature
  Communications}\ }\textbf {\bibinfo {volume} {4}},\ \bibinfo {pages} {1743}
  (\bibinfo {year} {2013})}\BibitemShut {NoStop}%
\bibitem [{\citenamefont {Tyryshkin}\ \emph {et~al.}(2012)\citenamefont
  {Tyryshkin}, \citenamefont {Tojo}, \citenamefont {Morton}, \citenamefont
  {Riemann}, \citenamefont {Abrosimov}, \citenamefont {Becker}, \citenamefont
  {Pohl}, \citenamefont {Schenkel}, \citenamefont {Thewalt}, \citenamefont
  {Itoh},\ and\ \citenamefont {Lyon}}]{tyryshkin_electron_2012}%
  \BibitemOpen
  \bibfield  {author} {\bibinfo {author} {\bibfnamefont {A.}~\bibnamefont
  {Tyryshkin}}, \bibinfo {author} {\bibfnamefont {S.}~\bibnamefont {Tojo}},
  \bibinfo {author} {\bibfnamefont {J.}~\bibnamefont {Morton}}, \bibinfo
  {author} {\bibfnamefont {H.}~\bibnamefont {Riemann}}, \bibinfo {author}
  {\bibfnamefont {N.}~\bibnamefont {Abrosimov}}, \bibinfo {author}
  {\bibfnamefont {P.}~\bibnamefont {Becker}}, \bibinfo {author} {\bibfnamefont
  {H.-J.}\ \bibnamefont {Pohl}}, \bibinfo {author} {\bibfnamefont
  {T.}~\bibnamefont {Schenkel}}, \bibinfo {author} {\bibfnamefont
  {M.}~\bibnamefont {Thewalt}}, \bibinfo {author} {\bibfnamefont
  {K.}~\bibnamefont {Itoh}}, \ and\ \bibinfo {author} {\bibfnamefont
  {S.}~\bibnamefont {Lyon}},\ }\href {\doibase 10.1038/nmat3182} {\bibfield
  {journal} {\bibinfo  {journal} {Nature Materials}\ }\textbf {\bibinfo
  {volume} {11}},\ \bibinfo {pages} {143} (\bibinfo {year} {2012})}\BibitemShut
  {NoStop}%
\bibitem [{\citenamefont {Wolfowicz}\ \emph {et~al.}(2013)\citenamefont
  {Wolfowicz}, \citenamefont {Tyryshkin}, \citenamefont {George}, \citenamefont
  {Riemann}, \citenamefont {Abrosimov}, \citenamefont {Becker}, \citenamefont
  {Pohl}, \citenamefont {Thewalt}, \citenamefont {Lyon},\ and\ \citenamefont
  {Morton}}]{wolfowicz_atomic_2013}%
  \BibitemOpen
  \bibfield  {author} {\bibinfo {author} {\bibfnamefont {G.}~\bibnamefont
  {Wolfowicz}}, \bibinfo {author} {\bibfnamefont {A.~M.}\ \bibnamefont
  {Tyryshkin}}, \bibinfo {author} {\bibfnamefont {R.~E.}\ \bibnamefont
  {George}}, \bibinfo {author} {\bibfnamefont {H.}~\bibnamefont {Riemann}},
  \bibinfo {author} {\bibfnamefont {N.~V.}\ \bibnamefont {Abrosimov}}, \bibinfo
  {author} {\bibfnamefont {P.}~\bibnamefont {Becker}}, \bibinfo {author}
  {\bibfnamefont {H.-J.}\ \bibnamefont {Pohl}}, \bibinfo {author}
  {\bibfnamefont {M.~L.~W.}\ \bibnamefont {Thewalt}}, \bibinfo {author}
  {\bibfnamefont {S.~A.}\ \bibnamefont {Lyon}}, \ and\ \bibinfo {author}
  {\bibfnamefont {J.~J.~L.}\ \bibnamefont {Morton}},\ }\href {\doibase
  10.1038/nnano.2013.117} {\bibfield  {journal} {\bibinfo  {journal} {Nature
  Nanotechnology}\ }\textbf {\bibinfo {volume} {8}},\ \bibinfo {pages} {561}
  (\bibinfo {year} {2013})}\BibitemShut {NoStop}%
\bibitem [{\citenamefont {Steger}\ \emph {et~al.}(2012)\citenamefont {Steger},
  \citenamefont {Saeedi}, \citenamefont {Thewalt}, \citenamefont {Morton},
  \citenamefont {Riemann}, \citenamefont {Abrosimov}, \citenamefont {Becker},\
  and\ \citenamefont {Pohl}}]{steger_quantum_2012}%
  \BibitemOpen
  \bibfield  {author} {\bibinfo {author} {\bibfnamefont {M.}~\bibnamefont
  {Steger}}, \bibinfo {author} {\bibfnamefont {K.}~\bibnamefont {Saeedi}},
  \bibinfo {author} {\bibfnamefont {M.~L.~W.}\ \bibnamefont {Thewalt}},
  \bibinfo {author} {\bibfnamefont {J.~J.~L.}\ \bibnamefont {Morton}}, \bibinfo
  {author} {\bibfnamefont {H.}~\bibnamefont {Riemann}}, \bibinfo {author}
  {\bibfnamefont {N.~V.}\ \bibnamefont {Abrosimov}}, \bibinfo {author}
  {\bibfnamefont {P.}~\bibnamefont {Becker}}, \ and\ \bibinfo {author}
  {\bibfnamefont {H.-J.}\ \bibnamefont {Pohl}},\ }\href {\doibase
  10.1126/science.1217635} {\bibfield  {journal} {\bibinfo  {journal}
  {Science}\ }\textbf {\bibinfo {volume} {336}},\ \bibinfo {pages} {1280}
  (\bibinfo {year} {2012})}\BibitemShut {NoStop}%
\bibitem [{\citenamefont {Saeedi}\ \emph {et~al.}(2013)\citenamefont {Saeedi},
  \citenamefont {Simmons}, \citenamefont {Salvail}, \citenamefont {Dluhy},
  \citenamefont {Riemann}, \citenamefont {Abrosimov}, \citenamefont {Becker},
  \citenamefont {Pohl}, \citenamefont {Morton},\ and\ \citenamefont
  {Thewalt}}]{Saeedi2013}%
  \BibitemOpen
  \bibfield  {author} {\bibinfo {author} {\bibfnamefont {K.}~\bibnamefont
  {Saeedi}}, \bibinfo {author} {\bibfnamefont {S.}~\bibnamefont {Simmons}},
  \bibinfo {author} {\bibfnamefont {J.~Z.}\ \bibnamefont {Salvail}}, \bibinfo
  {author} {\bibfnamefont {P.}~\bibnamefont {Dluhy}}, \bibinfo {author}
  {\bibfnamefont {H.}~\bibnamefont {Riemann}}, \bibinfo {author} {\bibfnamefont
  {N.~V.}\ \bibnamefont {Abrosimov}}, \bibinfo {author} {\bibfnamefont
  {P.}~\bibnamefont {Becker}}, \bibinfo {author} {\bibfnamefont {H.-J.}\
  \bibnamefont {Pohl}}, \bibinfo {author} {\bibfnamefont {J.~J.~L.}\
  \bibnamefont {Morton}}, \ and\ \bibinfo {author} {\bibfnamefont {M.~L.~W.}\
  \bibnamefont {Thewalt}},\ }\href {\doibase 10.1126/science.1239584}
  {\bibfield  {journal} {\bibinfo  {journal} {Science}\ }\textbf {\bibinfo
  {volume} {342}},\ \bibinfo {pages} {830} (\bibinfo {year}
  {2013})}\BibitemShut {NoStop}%
\bibitem [{\citenamefont {Zhong}\ \emph {et~al.}(2015)\citenamefont {Zhong},
  \citenamefont {Hedges}, \citenamefont {Ahlefeldt}, \citenamefont
  {Bartholomew}, \citenamefont {Beavan}, \citenamefont {Wittig}, \citenamefont
  {Longdell},\ and\ \citenamefont {Sellars}}]{Zhong.Nature.517.177(2015)}%
  \BibitemOpen
  \bibfield  {author} {\bibinfo {author} {\bibfnamefont {M.}~\bibnamefont
  {Zhong}}, \bibinfo {author} {\bibfnamefont {M.~P.}\ \bibnamefont {Hedges}},
  \bibinfo {author} {\bibfnamefont {R.~L.}\ \bibnamefont {Ahlefeldt}}, \bibinfo
  {author} {\bibfnamefont {J.~G.}\ \bibnamefont {Bartholomew}}, \bibinfo
  {author} {\bibfnamefont {S.~E.}\ \bibnamefont {Beavan}}, \bibinfo {author}
  {\bibfnamefont {S.~M.}\ \bibnamefont {Wittig}}, \bibinfo {author}
  {\bibfnamefont {J.~J.}\ \bibnamefont {Longdell}}, \ and\ \bibinfo {author}
  {\bibfnamefont {M.~J.}\ \bibnamefont {Sellars}},\ }\href@noop {} {\bibfield
  {journal} {\bibinfo  {journal} {Nature}\ }\textbf {\bibinfo {volume} {517}},\
  \bibinfo {pages} {177} (\bibinfo {year} {2015})}\BibitemShut {NoStop}%
\bibitem [{\citenamefont {Elzerman}\ \emph {et~al.}(2004)\citenamefont
  {Elzerman}, \citenamefont {Hanson}, \citenamefont {Willems~van Beveren},
  \citenamefont {Witkamp}, \citenamefont {Vandersypen},\ and\ \citenamefont
  {Kouwenhoven}}]{Elzerman.Nature.430.431(2004)}%
  \BibitemOpen
  \bibfield  {author} {\bibinfo {author} {\bibfnamefont {J.}~\bibnamefont
  {Elzerman}}, \bibinfo {author} {\bibfnamefont {R.}~\bibnamefont {Hanson}},
  \bibinfo {author} {\bibfnamefont {L.}~\bibnamefont {Willems~van Beveren}},
  \bibinfo {author} {\bibfnamefont {B.}~\bibnamefont {Witkamp}}, \bibinfo
  {author} {\bibfnamefont {L.}~\bibnamefont {Vandersypen}}, \ and\ \bibinfo
  {author} {\bibfnamefont {L.}~\bibnamefont {Kouwenhoven}},\ }\href@noop {}
  {\bibfield  {journal} {\bibinfo  {journal} {Nature}\ }\textbf {\bibinfo
  {volume} {430}},\ \bibinfo {pages} {431} (\bibinfo {year}
  {2004})}\BibitemShut {NoStop}%
\bibitem [{\citenamefont {Veldhorst}\ \emph {et~al.}(2014)\citenamefont
  {Veldhorst}, \citenamefont {Hwang}, \citenamefont {Yang}, \citenamefont
  {Leenstra}, \citenamefont {de~Ronde}, \citenamefont {Dehollain},
  \citenamefont {Muhonen}, \citenamefont {Hudson}, \citenamefont {Itoh},
  \citenamefont {Morello},\ and\ \citenamefont {Dzurak}}]{Veldhorst2014}%
  \BibitemOpen
  \bibfield  {author} {\bibinfo {author} {\bibfnamefont {M.}~\bibnamefont
  {Veldhorst}}, \bibinfo {author} {\bibfnamefont {J.~C.~C.}\ \bibnamefont
  {Hwang}}, \bibinfo {author} {\bibfnamefont {C.~H.}\ \bibnamefont {Yang}},
  \bibinfo {author} {\bibfnamefont {a.~W.}\ \bibnamefont {Leenstra}}, \bibinfo
  {author} {\bibfnamefont {B.}~\bibnamefont {de~Ronde}}, \bibinfo {author}
  {\bibfnamefont {J.~P.}\ \bibnamefont {Dehollain}}, \bibinfo {author}
  {\bibfnamefont {J.~T.}\ \bibnamefont {Muhonen}}, \bibinfo {author}
  {\bibfnamefont {F.~E.}\ \bibnamefont {Hudson}}, \bibinfo {author}
  {\bibfnamefont {K.~M.}\ \bibnamefont {Itoh}}, \bibinfo {author}
  {\bibfnamefont {A.}~\bibnamefont {Morello}}, \ and\ \bibinfo {author}
  {\bibfnamefont {a.~S.}\ \bibnamefont {Dzurak}},\ }\href {\doibase
  10.1038/nnano.2014.216} {\bibfield  {journal} {\bibinfo  {journal} {Nature
  Nanotechnology}\ }\textbf {\bibinfo {volume} {9}},\ \bibinfo {pages} {1}
  (\bibinfo {year} {2014})}\BibitemShut {NoStop}%
\bibitem [{\citenamefont {Petersson}\ \emph {et~al.}(2012)\citenamefont
  {Petersson}, \citenamefont {McFaul}, \citenamefont {Schroer}, \citenamefont
  {Jung}, \citenamefont {Taylor}, \citenamefont {Houck},\ and\ \citenamefont
  {Petta}}]{Petersson.Nature.490.380(2012)}%
  \BibitemOpen
  \bibfield  {author} {\bibinfo {author} {\bibfnamefont {K.}~\bibnamefont
  {Petersson}}, \bibinfo {author} {\bibfnamefont {L.}~\bibnamefont {McFaul}},
  \bibinfo {author} {\bibfnamefont {M.}~\bibnamefont {Schroer}}, \bibinfo
  {author} {\bibfnamefont {M.}~\bibnamefont {Jung}}, \bibinfo {author}
  {\bibfnamefont {J.}~\bibnamefont {Taylor}}, \bibinfo {author} {\bibfnamefont
  {A.}~\bibnamefont {Houck}}, \ and\ \bibinfo {author} {\bibfnamefont
  {J.}~\bibnamefont {Petta}},\ }\href@noop {} {\bibfield  {journal} {\bibinfo
  {journal} {Nature}\ }\textbf {\bibinfo {volume} {490}},\ \bibinfo {pages}
  {380} (\bibinfo {year} {2012})}\BibitemShut {NoStop}%
\bibitem [{\citenamefont {Morello}\ \emph {et~al.}(2010)\citenamefont
  {Morello}, \citenamefont {Pla}, \citenamefont {Zwanenburg}, \citenamefont
  {Chan}, \citenamefont {Tan}, \citenamefont {Huebl}, \citenamefont
  {Möttönen}, \citenamefont {Nugroho}, \citenamefont {Yang}, \citenamefont
  {van Donkelaar}, \citenamefont {Alves}, \citenamefont {Jamieson},
  \citenamefont {Escott}, \citenamefont {Hollenberg}, \citenamefont {Clark},\
  and\ \citenamefont {Dzurak}}]{morello_single-shot_2010}%
  \BibitemOpen
  \bibfield  {author} {\bibinfo {author} {\bibfnamefont {A.}~\bibnamefont
  {Morello}}, \bibinfo {author} {\bibfnamefont {J.~J.}\ \bibnamefont {Pla}},
  \bibinfo {author} {\bibfnamefont {F.~A.}\ \bibnamefont {Zwanenburg}},
  \bibinfo {author} {\bibfnamefont {K.~W.}\ \bibnamefont {Chan}}, \bibinfo
  {author} {\bibfnamefont {K.~Y.}\ \bibnamefont {Tan}}, \bibinfo {author}
  {\bibfnamefont {H.}~\bibnamefont {Huebl}}, \bibinfo {author} {\bibfnamefont
  {M.}~\bibnamefont {Möttönen}}, \bibinfo {author} {\bibfnamefont {C.~D.}\
  \bibnamefont {Nugroho}}, \bibinfo {author} {\bibfnamefont {C.}~\bibnamefont
  {Yang}}, \bibinfo {author} {\bibfnamefont {J.~A.}\ \bibnamefont {van
  Donkelaar}}, \bibinfo {author} {\bibfnamefont {A.~D.~C.}\ \bibnamefont
  {Alves}}, \bibinfo {author} {\bibfnamefont {D.~N.}\ \bibnamefont {Jamieson}},
  \bibinfo {author} {\bibfnamefont {C.~C.}\ \bibnamefont {Escott}}, \bibinfo
  {author} {\bibfnamefont {L.~C.~L.}\ \bibnamefont {Hollenberg}}, \bibinfo
  {author} {\bibfnamefont {R.~G.}\ \bibnamefont {Clark}}, \ and\ \bibinfo
  {author} {\bibfnamefont {A.~S.}\ \bibnamefont {Dzurak}},\ }\href {\doibase
  10.1038/nature09392} {\bibfield  {journal} {\bibinfo  {journal} {Nature}\
  }\textbf {\bibinfo {volume} {467}},\ \bibinfo {pages} {687} (\bibinfo {year}
  {2010})}\BibitemShut {NoStop}%
\bibitem [{\citenamefont {Pla}\ \emph {et~al.}(2012)\citenamefont {Pla},
  \citenamefont {Tan}, \citenamefont {Dehollain}, \citenamefont {Lim},
  \citenamefont {Morton}, \citenamefont {Jamieson}, \citenamefont {Dzurak},\
  and\ \citenamefont {Morello}}]{pla_single-atom_2012}%
  \BibitemOpen
  \bibfield  {author} {\bibinfo {author} {\bibfnamefont {J.~J.}\ \bibnamefont
  {Pla}}, \bibinfo {author} {\bibfnamefont {K.~Y.}\ \bibnamefont {Tan}},
  \bibinfo {author} {\bibfnamefont {J.~P.}\ \bibnamefont {Dehollain}}, \bibinfo
  {author} {\bibfnamefont {W.~H.}\ \bibnamefont {Lim}}, \bibinfo {author}
  {\bibfnamefont {J.~J.~L.}\ \bibnamefont {Morton}}, \bibinfo {author}
  {\bibfnamefont {D.~N.}\ \bibnamefont {Jamieson}}, \bibinfo {author}
  {\bibfnamefont {A.~S.}\ \bibnamefont {Dzurak}}, \ and\ \bibinfo {author}
  {\bibfnamefont {A.}~\bibnamefont {Morello}},\ }\href {\doibase
  10.1038/nature11449} {\bibfield  {journal} {\bibinfo  {journal} {Nature}\
  }\textbf {\bibinfo {volume} {489}},\ \bibinfo {pages} {541} (\bibinfo {year}
  {2012})}\BibitemShut {NoStop}%
\bibitem [{\citenamefont {Vincent}\ \emph {et~al.}(2012)\citenamefont
  {Vincent}, \citenamefont {Klyatskaya}, \citenamefont {Ruben}, \citenamefont
  {Wernsdorfer},\ and\ \citenamefont {Balestro}}]{vincent2012electronic}%
  \BibitemOpen
  \bibfield  {author} {\bibinfo {author} {\bibfnamefont {R.}~\bibnamefont
  {Vincent}}, \bibinfo {author} {\bibfnamefont {S.}~\bibnamefont {Klyatskaya}},
  \bibinfo {author} {\bibfnamefont {M.}~\bibnamefont {Ruben}}, \bibinfo
  {author} {\bibfnamefont {W.}~\bibnamefont {Wernsdorfer}}, \ and\ \bibinfo
  {author} {\bibfnamefont {F.}~\bibnamefont {Balestro}},\ }\href@noop {}
  {\bibfield  {journal} {\bibinfo  {journal} {Nature}\ }\textbf {\bibinfo
  {volume} {488}},\ \bibinfo {pages} {357} (\bibinfo {year}
  {2012})}\BibitemShut {NoStop}%
\bibitem [{\citenamefont {Thiele}\ \emph {et~al.}(2013)\citenamefont {Thiele},
  \citenamefont {Vincent}, \citenamefont {Holzmann}, \citenamefont
  {Klyatskaya}, \citenamefont {Ruben}, \citenamefont {Balestro},\ and\
  \citenamefont {Wernsdorfer}}]{Thiele.PhysRevLett.111.037203(2013)}%
  \BibitemOpen
  \bibfield  {author} {\bibinfo {author} {\bibfnamefont {S.}~\bibnamefont
  {Thiele}}, \bibinfo {author} {\bibfnamefont {R.}~\bibnamefont {Vincent}},
  \bibinfo {author} {\bibfnamefont {M.}~\bibnamefont {Holzmann}}, \bibinfo
  {author} {\bibfnamefont {S.}~\bibnamefont {Klyatskaya}}, \bibinfo {author}
  {\bibfnamefont {M.}~\bibnamefont {Ruben}}, \bibinfo {author} {\bibfnamefont
  {F.}~\bibnamefont {Balestro}}, \ and\ \bibinfo {author} {\bibfnamefont
  {W.}~\bibnamefont {Wernsdorfer}},\ }\href {\doibase
  10.1103/PhysRevLett.111.037203} {\bibfield  {journal} {\bibinfo  {journal}
  {Phys. Rev. Lett.}\ }\textbf {\bibinfo {volume} {111}},\ \bibinfo {pages}
  {037203} (\bibinfo {year} {2013})}\BibitemShut {NoStop}%
\bibitem [{\citenamefont {K{\"o}hler}\ \emph {et~al.}(1993)\citenamefont
  {K{\"o}hler}, \citenamefont {Disselhorst}, \citenamefont {Donckers},
  \citenamefont {Groenen}, \citenamefont {Schmidt},\ and\ \citenamefont
  {Moerner}}]{Kohler.Nature.363.242(1993)}%
  \BibitemOpen
  \bibfield  {author} {\bibinfo {author} {\bibfnamefont {J.}~\bibnamefont
  {K{\"o}hler}}, \bibinfo {author} {\bibfnamefont {J.}~\bibnamefont
  {Disselhorst}}, \bibinfo {author} {\bibfnamefont {M.}~\bibnamefont
  {Donckers}}, \bibinfo {author} {\bibfnamefont {E.}~\bibnamefont {Groenen}},
  \bibinfo {author} {\bibfnamefont {J.}~\bibnamefont {Schmidt}}, \ and\
  \bibinfo {author} {\bibfnamefont {W.}~\bibnamefont {Moerner}},\ }\href@noop
  {} {\bibfield  {journal} {\bibinfo  {journal} {Nature}\ }\textbf {\bibinfo
  {volume} {363}},\ \bibinfo {pages} {242} (\bibinfo {year}
  {1993})}\BibitemShut {NoStop}%
\bibitem [{\citenamefont {Wrachtrup}\ \emph {et~al.}(1993)\citenamefont
  {Wrachtrup}, \citenamefont {Von~Borczyskowski}, \citenamefont {Bernard},
  \citenamefont {Orritt},\ and\ \citenamefont
  {Brown}}]{Wrachtrup.Nature.363.244(1993)}%
  \BibitemOpen
  \bibfield  {author} {\bibinfo {author} {\bibfnamefont {J.}~\bibnamefont
  {Wrachtrup}}, \bibinfo {author} {\bibfnamefont {C.}~\bibnamefont
  {Von~Borczyskowski}}, \bibinfo {author} {\bibfnamefont {J.}~\bibnamefont
  {Bernard}}, \bibinfo {author} {\bibfnamefont {M.}~\bibnamefont {Orritt}}, \
  and\ \bibinfo {author} {\bibfnamefont {R.}~\bibnamefont {Brown}},\
  }\href@noop {} {\bibfield  {journal} {\bibinfo  {journal} {Nature}\ }\textbf
  {\bibinfo {volume} {363}},\ \bibinfo {pages} {244} (\bibinfo {year}
  {1993})}\BibitemShut {NoStop}%
\bibitem [{\citenamefont {Jelezko}\ \emph {et~al.}(2004)\citenamefont
  {Jelezko}, \citenamefont {Gaebel}, \citenamefont {Popa}, \citenamefont
  {Gruber},\ and\ \citenamefont {Wrachtrup}}]{jelezko_observation_2004}%
  \BibitemOpen
  \bibfield  {author} {\bibinfo {author} {\bibfnamefont {F.}~\bibnamefont
  {Jelezko}}, \bibinfo {author} {\bibfnamefont {T.}~\bibnamefont {Gaebel}},
  \bibinfo {author} {\bibfnamefont {I.}~\bibnamefont {Popa}}, \bibinfo {author}
  {\bibfnamefont {A.}~\bibnamefont {Gruber}}, \ and\ \bibinfo {author}
  {\bibfnamefont {J.}~\bibnamefont {Wrachtrup}},\ }\href {\doibase
  10.1103/PhysRevLett.92.076401} {\bibfield  {journal} {\bibinfo  {journal}
  {Phys. Rev. Lett.}\ }\textbf {\bibinfo {volume} {92}},\ \bibinfo {pages}
  {076401} (\bibinfo {year} {2004})}\BibitemShut {NoStop}%
\bibitem [{\citenamefont {Manson}\ \emph {et~al.}(2006)\citenamefont {Manson},
  \citenamefont {Harrison},\ and\ \citenamefont
  {Sellars}}]{Manson.PhysRevB.74.104303(2006)}%
  \BibitemOpen
  \bibfield  {author} {\bibinfo {author} {\bibfnamefont {N.~B.}\ \bibnamefont
  {Manson}}, \bibinfo {author} {\bibfnamefont {J.~P.}\ \bibnamefont
  {Harrison}}, \ and\ \bibinfo {author} {\bibfnamefont {M.~J.}\ \bibnamefont
  {Sellars}},\ }\href {\doibase 10.1103/PhysRevB.74.104303} {\bibfield
  {journal} {\bibinfo  {journal} {Phys. Rev. B}\ }\textbf {\bibinfo {volume}
  {74}},\ \bibinfo {pages} {104303} (\bibinfo {year} {2006})}\BibitemShut
  {NoStop}%
\bibitem [{\citenamefont {Widmann}\ \emph {et~al.}(2015)\citenamefont
  {Widmann}, \citenamefont {Lee}, \citenamefont {Rendler}, \citenamefont {Son},
  \citenamefont {Fedder}, \citenamefont {Paik}, \citenamefont {Yang},
  \citenamefont {Zhao}, \citenamefont {Yang}, \citenamefont {Booker} \emph
  {et~al.}}]{Widmann.NatureMat.14.164(2015)}%
  \BibitemOpen
  \bibfield  {author} {\bibinfo {author} {\bibfnamefont {M.}~\bibnamefont
  {Widmann}}, \bibinfo {author} {\bibfnamefont {S.-Y.}\ \bibnamefont {Lee}},
  \bibinfo {author} {\bibfnamefont {T.}~\bibnamefont {Rendler}}, \bibinfo
  {author} {\bibfnamefont {N.~T.}\ \bibnamefont {Son}}, \bibinfo {author}
  {\bibfnamefont {H.}~\bibnamefont {Fedder}}, \bibinfo {author} {\bibfnamefont
  {S.}~\bibnamefont {Paik}}, \bibinfo {author} {\bibfnamefont {L.-P.}\
  \bibnamefont {Yang}}, \bibinfo {author} {\bibfnamefont {N.}~\bibnamefont
  {Zhao}}, \bibinfo {author} {\bibfnamefont {S.}~\bibnamefont {Yang}}, \bibinfo
  {author} {\bibfnamefont {I.}~\bibnamefont {Booker}},  \emph {et~al.},\
  }\href@noop {} {\bibfield  {journal} {\bibinfo  {journal} {Nature materials}\
  }\textbf {\bibinfo {volume} {14}},\ \bibinfo {pages} {164} (\bibinfo {year}
  {2015})}\BibitemShut {NoStop}%
\bibitem [{\citenamefont {Rugar}\ \emph {et~al.}(2004)\citenamefont {Rugar},
  \citenamefont {Budakian}, \citenamefont {Mamin},\ and\ \citenamefont
  {Chui}}]{Rugar.Nature.430.329(2004)}%
  \BibitemOpen
  \bibfield  {author} {\bibinfo {author} {\bibfnamefont {D.}~\bibnamefont
  {Rugar}}, \bibinfo {author} {\bibfnamefont {R.}~\bibnamefont {Budakian}},
  \bibinfo {author} {\bibfnamefont {H.}~\bibnamefont {Mamin}}, \ and\ \bibinfo
  {author} {\bibfnamefont {B.}~\bibnamefont {Chui}},\ }\href@noop {} {\bibfield
   {journal} {\bibinfo  {journal} {Nature}\ }\textbf {\bibinfo {volume}
  {430}},\ \bibinfo {pages} {329} (\bibinfo {year} {2004})}\BibitemShut
  {NoStop}%
\bibitem [{\citenamefont {Grinolds}\ \emph {et~al.}(2013)\citenamefont
  {Grinolds}, \citenamefont {Hong}, \citenamefont {Maletinsky}, \citenamefont
  {Luan}, \citenamefont {Lukin}, \citenamefont {Walsworth},\ and\ \citenamefont
  {Yacoby}}]{Grinolds.NaturePhys.9.215(2013)}%
  \BibitemOpen
  \bibfield  {author} {\bibinfo {author} {\bibfnamefont {M.~S.}\ \bibnamefont
  {Grinolds}}, \bibinfo {author} {\bibfnamefont {S.}~\bibnamefont {Hong}},
  \bibinfo {author} {\bibfnamefont {P.}~\bibnamefont {Maletinsky}}, \bibinfo
  {author} {\bibfnamefont {L.}~\bibnamefont {Luan}}, \bibinfo {author}
  {\bibfnamefont {M.~D.}\ \bibnamefont {Lukin}}, \bibinfo {author}
  {\bibfnamefont {R.~L.}\ \bibnamefont {Walsworth}}, \ and\ \bibinfo {author}
  {\bibfnamefont {A.}~\bibnamefont {Yacoby}},\ }\href@noop {} {\bibfield
  {journal} {\bibinfo  {journal} {Nature Physics}\ }\textbf {\bibinfo {volume}
  {9}},\ \bibinfo {pages} {215} (\bibinfo {year} {2013})}\BibitemShut {NoStop}%
\bibitem [{\citenamefont {Manassen}\ \emph {et~al.}(1989)\citenamefont
  {Manassen}, \citenamefont {Hamers}, \citenamefont {Demuth},\ and\
  \citenamefont {Castellano~Jr.}}]{Manassen.PhysRevLett.62.2531(1989)}%
  \BibitemOpen
  \bibfield  {author} {\bibinfo {author} {\bibfnamefont {Y.}~\bibnamefont
  {Manassen}}, \bibinfo {author} {\bibfnamefont {R.~J.}\ \bibnamefont
  {Hamers}}, \bibinfo {author} {\bibfnamefont {J.~E.}\ \bibnamefont {Demuth}},
  \ and\ \bibinfo {author} {\bibfnamefont {A.~J.}\ \bibnamefont
  {Castellano~Jr.}},\ }\href {\doibase 10.1103/PhysRevLett.62.2531} {\bibfield
  {journal} {\bibinfo  {journal} {Phys. Rev. Lett.}\ }\textbf {\bibinfo
  {volume} {62}},\ \bibinfo {pages} {2531} (\bibinfo {year}
  {1989})}\BibitemShut {NoStop}%
\bibitem [{\citenamefont {Baumann}\ \emph {et~al.}(2015)\citenamefont
  {Baumann}, \citenamefont {Paul}, \citenamefont {Choi}, \citenamefont {Lutz},
  \citenamefont {Ardavan},\ and\ \citenamefont
  {Heinrich}}]{baumann2015electron}%
  \BibitemOpen
  \bibfield  {author} {\bibinfo {author} {\bibfnamefont {S.}~\bibnamefont
  {Baumann}}, \bibinfo {author} {\bibfnamefont {W.}~\bibnamefont {Paul}},
  \bibinfo {author} {\bibfnamefont {T.}~\bibnamefont {Choi}}, \bibinfo {author}
  {\bibfnamefont {C.~P.}\ \bibnamefont {Lutz}}, \bibinfo {author}
  {\bibfnamefont {A.}~\bibnamefont {Ardavan}}, \ and\ \bibinfo {author}
  {\bibfnamefont {A.~J.}\ \bibnamefont {Heinrich}},\ }\href@noop {} {\bibfield
  {journal} {\bibinfo  {journal} {Science}\ }\textbf {\bibinfo {volume}
  {350}},\ \bibinfo {pages} {417} (\bibinfo {year} {2015})}\BibitemShut
  {NoStop}%
\bibitem [{\citenamefont {Schweiger}\ and\ \citenamefont
  {Jeschke}(2001)}]{SchweigerEPR(2001)}%
  \BibitemOpen
  \bibfield  {author} {\bibinfo {author} {\bibfnamefont {A.}~\bibnamefont
  {Schweiger}}\ and\ \bibinfo {author} {\bibfnamefont {G.}~\bibnamefont
  {Jeschke}},\ }\href@noop {} {\emph {\bibinfo {title} {Principles of pulse
  electron paramagnetic resonance}}}\ (\bibinfo  {publisher} {Oxford University
  Press},\ \bibinfo {year} {2001})\BibitemShut {NoStop}%
\bibitem [{\citenamefont {Sigillito}\ \emph {et~al.}(2014)\citenamefont
  {Sigillito}, \citenamefont {Malissa}, \citenamefont {Tyryshkin},
  \citenamefont {Riemann}, \citenamefont {Abrosimov}, \citenamefont {Becker},
  \citenamefont {Pohl}, \citenamefont {Thewalt}, \citenamefont {Itoh},
  \citenamefont {Morton}, \citenamefont {Houck}, \citenamefont {Schuster},\
  and\ \citenamefont {Lyon}}]{Sigillito.APL.104.104.22407(2014)}%
  \BibitemOpen
  \bibfield  {author} {\bibinfo {author} {\bibfnamefont {A.~J.}\ \bibnamefont
  {Sigillito}}, \bibinfo {author} {\bibfnamefont {H.}~\bibnamefont {Malissa}},
  \bibinfo {author} {\bibfnamefont {A.~M.}\ \bibnamefont {Tyryshkin}}, \bibinfo
  {author} {\bibfnamefont {H.}~\bibnamefont {Riemann}}, \bibinfo {author}
  {\bibfnamefont {N.~V.}\ \bibnamefont {Abrosimov}}, \bibinfo {author}
  {\bibfnamefont {P.}~\bibnamefont {Becker}}, \bibinfo {author} {\bibfnamefont
  {H.-J.}\ \bibnamefont {Pohl}}, \bibinfo {author} {\bibfnamefont {M.~L.~W.}\
  \bibnamefont {Thewalt}}, \bibinfo {author} {\bibfnamefont {K.~M.}\
  \bibnamefont {Itoh}}, \bibinfo {author} {\bibfnamefont {J.~J.~L.}\
  \bibnamefont {Morton}}, \bibinfo {author} {\bibfnamefont {A.~A.}\
  \bibnamefont {Houck}}, \bibinfo {author} {\bibfnamefont {D.~I.}\ \bibnamefont
  {Schuster}}, \ and\ \bibinfo {author} {\bibfnamefont {S.~A.}\ \bibnamefont
  {Lyon}},\ }\href {\doibase http://dx.doi.org/10.1063/1.4881613} {\bibfield
  {journal} {\bibinfo  {journal} {Applied Physics Letters}\ }\textbf {\bibinfo
  {volume} {104}},\ \bibinfo {eid} {222407} (\bibinfo {year}
  {2014})}\BibitemShut {NoStop}%
\bibitem [{\citenamefont {Bienfait}\ \emph
  {et~al.}(2016{\natexlab{a}})\citenamefont {Bienfait}, \citenamefont {Pla},
  \citenamefont {Kubo}, \citenamefont {Stern}, \citenamefont {Zhou},
  \citenamefont {Lo}, \citenamefont {Weis}, \citenamefont {Schenkel},
  \citenamefont {Thewalt}, \citenamefont {Vion}, \citenamefont {Esteve},
  \citenamefont {Julsgaard}, \citenamefont {Moelmer}, \citenamefont {Morton},\
  and\ \citenamefont {Bertet}}]{bienfait2015reaching}%
  \BibitemOpen
  \bibfield  {author} {\bibinfo {author} {\bibfnamefont {A.}~\bibnamefont
  {Bienfait}}, \bibinfo {author} {\bibfnamefont {J.}~\bibnamefont {Pla}},
  \bibinfo {author} {\bibfnamefont {Y.}~\bibnamefont {Kubo}}, \bibinfo {author}
  {\bibfnamefont {M.}~\bibnamefont {Stern}}, \bibinfo {author} {\bibfnamefont
  {X.}~\bibnamefont {Zhou}}, \bibinfo {author} {\bibfnamefont {C.}~\bibnamefont
  {Lo}}, \bibinfo {author} {\bibfnamefont {C.}~\bibnamefont {Weis}}, \bibinfo
  {author} {\bibfnamefont {T.}~\bibnamefont {Schenkel}}, \bibinfo {author}
  {\bibfnamefont {M.}~\bibnamefont {Thewalt}}, \bibinfo {author} {\bibfnamefont
  {D.}~\bibnamefont {Vion}}, \bibinfo {author} {\bibfnamefont {D.}~\bibnamefont
  {Esteve}}, \bibinfo {author} {\bibfnamefont {B.}~\bibnamefont {Julsgaard}},
  \bibinfo {author} {\bibfnamefont {K.}~\bibnamefont {Moelmer}}, \bibinfo
  {author} {\bibfnamefont {J.}~\bibnamefont {Morton}}, \ and\ \bibinfo {author}
  {\bibfnamefont {P.}~\bibnamefont {Bertet}},\ }\href@noop {} {\bibfield
  {journal} {\bibinfo  {journal} {Nature nanotechnology}\ }\textbf {\bibinfo
  {volume} {11}},\ \bibinfo {pages} {253} (\bibinfo {year}
  {2016}{\natexlab{a}})}\BibitemShut {NoStop}%
\bibitem [{\citenamefont {Castellanos-Beltran}\ and\ \citenamefont
  {Lehnert}(2007)}]{Castellanos-Beltran.ApplPhysLett.91.083509(2007)}%
  \BibitemOpen
  \bibfield  {author} {\bibinfo {author} {\bibfnamefont {M.~A.}\ \bibnamefont
  {Castellanos-Beltran}}\ and\ \bibinfo {author} {\bibfnamefont {K.~W.}\
  \bibnamefont {Lehnert}},\ }\href {\doibase 10.1063/1.2773988} {\bibfield
  {journal} {\bibinfo  {journal} {Appl. Phys. Lett.}\ }\textbf {\bibinfo
  {volume} {91}},\ \bibinfo {pages} {083509} (\bibinfo {year}
  {2007})}\BibitemShut {NoStop}%
\bibitem [{\citenamefont {Bergeal}\ \emph {et~al.}(2010)\citenamefont
  {Bergeal}, \citenamefont {Schackert}, \citenamefont {Metcalfe}, \citenamefont
  {Vijay}, \citenamefont {Manucharyan}, \citenamefont {Frunzio}, \citenamefont
  {Prober}, \citenamefont {Schoelkopf}, \citenamefont {Girvin},\ and\
  \citenamefont {Devoret}}]{bergeal_phase-preserving_2010}%
  \BibitemOpen
  \bibfield  {author} {\bibinfo {author} {\bibfnamefont {N.}~\bibnamefont
  {Bergeal}}, \bibinfo {author} {\bibfnamefont {F.}~\bibnamefont {Schackert}},
  \bibinfo {author} {\bibfnamefont {M.}~\bibnamefont {Metcalfe}}, \bibinfo
  {author} {\bibfnamefont {R.}~\bibnamefont {Vijay}}, \bibinfo {author}
  {\bibfnamefont {V.~E.}\ \bibnamefont {Manucharyan}}, \bibinfo {author}
  {\bibfnamefont {L.}~\bibnamefont {Frunzio}}, \bibinfo {author} {\bibfnamefont
  {D.~E.}\ \bibnamefont {Prober}}, \bibinfo {author} {\bibfnamefont {R.~J.}\
  \bibnamefont {Schoelkopf}}, \bibinfo {author} {\bibfnamefont {S.~M.}\
  \bibnamefont {Girvin}}, \ and\ \bibinfo {author} {\bibfnamefont {M.~H.}\
  \bibnamefont {Devoret}},\ }\href {\doibase 10.1038/nature09035} {\bibfield
  {journal} {\bibinfo  {journal} {Nature}\ }\textbf {\bibinfo {volume} {465}},\
  \bibinfo {pages} {64} (\bibinfo {year} {2010})}\BibitemShut {NoStop}%
\bibitem [{\citenamefont {Zhou}\ \emph {et~al.}(2014)\citenamefont {Zhou},
  \citenamefont {Schmitt}, \citenamefont {Bertet}, \citenamefont {Vion},
  \citenamefont {Wustmann}, \citenamefont {Shumeiko},\ and\ \citenamefont
  {Esteve}}]{Zhou.PhysRevB.89.214517(2014)}%
  \BibitemOpen
  \bibfield  {author} {\bibinfo {author} {\bibfnamefont {X.}~\bibnamefont
  {Zhou}}, \bibinfo {author} {\bibfnamefont {V.}~\bibnamefont {Schmitt}},
  \bibinfo {author} {\bibfnamefont {P.}~\bibnamefont {Bertet}}, \bibinfo
  {author} {\bibfnamefont {D.}~\bibnamefont {Vion}}, \bibinfo {author}
  {\bibfnamefont {W.}~\bibnamefont {Wustmann}}, \bibinfo {author}
  {\bibfnamefont {V.}~\bibnamefont {Shumeiko}}, \ and\ \bibinfo {author}
  {\bibfnamefont {D.}~\bibnamefont {Esteve}},\ }\href {\doibase
  10.1103/PhysRevB.89.214517} {\bibfield  {journal} {\bibinfo  {journal} {Phys.
  Rev. B}\ }\textbf {\bibinfo {volume} {89}},\ \bibinfo {pages} {214517}
  (\bibinfo {year} {2014})}\BibitemShut {NoStop}%
\bibitem [{\citenamefont {Eichler}\ \emph {et~al.}(2016)\citenamefont
  {Eichler}, \citenamefont {Sigillito}, \citenamefont {Lyon},\ and\
  \citenamefont {Petta}}]{eichler2016electron}%
  \BibitemOpen
  \bibfield  {author} {\bibinfo {author} {\bibfnamefont {C.}~\bibnamefont
  {Eichler}}, \bibinfo {author} {\bibfnamefont {A.}~\bibnamefont {Sigillito}},
  \bibinfo {author} {\bibfnamefont {S.}~\bibnamefont {Lyon}}, \ and\ \bibinfo
  {author} {\bibfnamefont {J.}~\bibnamefont {Petta}},\ }\href@noop {}
  {\bibfield  {journal} {\bibinfo  {journal} {arXiv preprint arXiv:1608.05130}\
  } (\bibinfo {year} {2016})}\BibitemShut {NoStop}%
\bibitem [{\citenamefont {Cottet}\ and\ \citenamefont
  {Kontos}(2010)}]{Cottet.PhysRevLett.105.160502}%
  \BibitemOpen
  \bibfield  {author} {\bibinfo {author} {\bibfnamefont {A.}~\bibnamefont
  {Cottet}}\ and\ \bibinfo {author} {\bibfnamefont {T.}~\bibnamefont
  {Kontos}},\ }\href {\doibase 10.1103/PhysRevLett.105.160502} {\bibfield
  {journal} {\bibinfo  {journal} {Phys. Rev. Lett.}\ }\textbf {\bibinfo
  {volume} {105}},\ \bibinfo {pages} {160502} (\bibinfo {year}
  {2010})}\BibitemShut {NoStop}%
\bibitem [{\citenamefont {Tosi}\ \emph {et~al.}(2015)\citenamefont {Tosi},
  \citenamefont {Mohiyaddin}, \citenamefont {Tenberg}, \citenamefont {Rahman},
  \citenamefont {Klimeck},\ and\ \citenamefont {Morello}}]{tosi2015silicon}%
  \BibitemOpen
  \bibfield  {author} {\bibinfo {author} {\bibfnamefont {G.}~\bibnamefont
  {Tosi}}, \bibinfo {author} {\bibfnamefont {F.~A.}\ \bibnamefont
  {Mohiyaddin}}, \bibinfo {author} {\bibfnamefont {S.~B.}\ \bibnamefont
  {Tenberg}}, \bibinfo {author} {\bibfnamefont {R.}~\bibnamefont {Rahman}},
  \bibinfo {author} {\bibfnamefont {G.}~\bibnamefont {Klimeck}}, \ and\
  \bibinfo {author} {\bibfnamefont {A.}~\bibnamefont {Morello}},\ }\href@noop
  {} {\bibfield  {journal} {\bibinfo  {journal} {arXiv preprint
  arXiv:1509.08538}\ } (\bibinfo {year} {2015})}\BibitemShut {NoStop}%
\bibitem [{\citenamefont {Beaudoin}\ \emph {et~al.}(2016)\citenamefont
  {Beaudoin}, \citenamefont {Lachance-Quirion}, \citenamefont {Coish},\ and\
  \citenamefont {Pioro-Ladri{\`e}re}}]{beaudoin2016coupling}%
  \BibitemOpen
  \bibfield  {author} {\bibinfo {author} {\bibfnamefont {F.}~\bibnamefont
  {Beaudoin}}, \bibinfo {author} {\bibfnamefont {D.}~\bibnamefont
  {Lachance-Quirion}}, \bibinfo {author} {\bibfnamefont {W.}~\bibnamefont
  {Coish}}, \ and\ \bibinfo {author} {\bibfnamefont {M.}~\bibnamefont
  {Pioro-Ladri{\`e}re}},\ }\href@noop {} {\bibfield  {journal} {\bibinfo
  {journal} {Nanotechnology}\ }\textbf {\bibinfo {volume} {27}},\ \bibinfo
  {pages} {464003} (\bibinfo {year} {2016})}\BibitemShut {NoStop}%
\bibitem [{\citenamefont {Samkharadze}\ \emph {et~al.}(2016)\citenamefont
  {Samkharadze}, \citenamefont {Bruno}, \citenamefont {Scarlino}, \citenamefont
  {Zheng}, \citenamefont {DiVincenzo}, \citenamefont {DiCarlo},\ and\
  \citenamefont {Vandersypen}}]{Samkharadze.PhysRevApplied.5.044004}%
  \BibitemOpen
  \bibfield  {author} {\bibinfo {author} {\bibfnamefont {N.}~\bibnamefont
  {Samkharadze}}, \bibinfo {author} {\bibfnamefont {A.}~\bibnamefont {Bruno}},
  \bibinfo {author} {\bibfnamefont {P.}~\bibnamefont {Scarlino}}, \bibinfo
  {author} {\bibfnamefont {G.}~\bibnamefont {Zheng}}, \bibinfo {author}
  {\bibfnamefont {D.~P.}\ \bibnamefont {DiVincenzo}}, \bibinfo {author}
  {\bibfnamefont {L.}~\bibnamefont {DiCarlo}}, \ and\ \bibinfo {author}
  {\bibfnamefont {L.~M.~K.}\ \bibnamefont {Vandersypen}},\ }\href {\doibase
  10.1103/PhysRevApplied.5.044004} {\bibfield  {journal} {\bibinfo  {journal}
  {Phys. Rev. Applied}\ }\textbf {\bibinfo {volume} {5}},\ \bibinfo {pages}
  {044004} (\bibinfo {year} {2016})}\BibitemShut {NoStop}%
\bibitem [{\citenamefont {Viennot}\ \emph {et~al.}(2015)\citenamefont
  {Viennot}, \citenamefont {Dartiailh}, \citenamefont {Cottet},\ and\
  \citenamefont {Kontos}}]{Viennot.Science.349.408(2015)}%
  \BibitemOpen
  \bibfield  {author} {\bibinfo {author} {\bibfnamefont {J.~J.}\ \bibnamefont
  {Viennot}}, \bibinfo {author} {\bibfnamefont {M.~C.}\ \bibnamefont
  {Dartiailh}}, \bibinfo {author} {\bibfnamefont {A.}~\bibnamefont {Cottet}}, \
  and\ \bibinfo {author} {\bibfnamefont {T.}~\bibnamefont {Kontos}},\
  }\href@noop {} {\bibfield  {journal} {\bibinfo  {journal} {Science}\ }\textbf
  {\bibinfo {volume} {349}},\ \bibinfo {pages} {408} (\bibinfo {year}
  {2015})}\BibitemShut {NoStop}%
\bibitem [{\citenamefont {Tosi}\ \emph {et~al.}(2014)\citenamefont {Tosi},
  \citenamefont {Mohiyaddin}, \citenamefont {Huebl},\ and\ \citenamefont
  {Morello}}]{Tosi.AIPAdv.4.087122(2014)}%
  \BibitemOpen
  \bibfield  {author} {\bibinfo {author} {\bibfnamefont {G.}~\bibnamefont
  {Tosi}}, \bibinfo {author} {\bibfnamefont {F.~A.}\ \bibnamefont
  {Mohiyaddin}}, \bibinfo {author} {\bibfnamefont {H.}~\bibnamefont {Huebl}}, \
  and\ \bibinfo {author} {\bibfnamefont {A.}~\bibnamefont {Morello}},\
  }\href@noop {} {\bibfield  {journal} {\bibinfo  {journal} {AIP Advances}\
  }\textbf {\bibinfo {volume} {4}},\ \bibinfo {pages} {087122} (\bibinfo {year}
  {2014})}\BibitemShut {NoStop}%
\bibitem [{\citenamefont {Campagne-Ibarcq}\ \emph {et~al.}(2013)\citenamefont
  {Campagne-Ibarcq}, \citenamefont {Flurin}, \citenamefont {Roch},
  \citenamefont {Darson}, \citenamefont {Morfin}, \citenamefont {Mirrahimi},
  \citenamefont {Devoret}, \citenamefont {Mallet},\ and\ \citenamefont
  {Huard}}]{Campagne.PhysRevX.3.021008(2013)}%
  \BibitemOpen
  \bibfield  {author} {\bibinfo {author} {\bibfnamefont {P.}~\bibnamefont
  {Campagne-Ibarcq}}, \bibinfo {author} {\bibfnamefont {E.}~\bibnamefont
  {Flurin}}, \bibinfo {author} {\bibfnamefont {N.}~\bibnamefont {Roch}},
  \bibinfo {author} {\bibfnamefont {D.}~\bibnamefont {Darson}}, \bibinfo
  {author} {\bibfnamefont {P.}~\bibnamefont {Morfin}}, \bibinfo {author}
  {\bibfnamefont {M.}~\bibnamefont {Mirrahimi}}, \bibinfo {author}
  {\bibfnamefont {M.~H.}\ \bibnamefont {Devoret}}, \bibinfo {author}
  {\bibfnamefont {F.}~\bibnamefont {Mallet}}, \ and\ \bibinfo {author}
  {\bibfnamefont {B.}~\bibnamefont {Huard}},\ }\href {\doibase
  10.1103/PhysRevX.3.021008} {\bibfield  {journal} {\bibinfo  {journal} {Phys.
  Rev. X}\ }\textbf {\bibinfo {volume} {3}},\ \bibinfo {pages} {021008}
  (\bibinfo {year} {2013})}\BibitemShut {NoStop}%
\bibitem [{\citenamefont {Jarmola}\ \emph {et~al.}(2012)\citenamefont
  {Jarmola}, \citenamefont {Acosta}, \citenamefont {Jensen}, \citenamefont
  {Chemerisov},\ and\ \citenamefont {Budker}}]{jarmola_temperature-_2012}%
  \BibitemOpen
  \bibfield  {author} {\bibinfo {author} {\bibfnamefont {A.}~\bibnamefont
  {Jarmola}}, \bibinfo {author} {\bibfnamefont {V.~M.}\ \bibnamefont {Acosta}},
  \bibinfo {author} {\bibfnamefont {K.}~\bibnamefont {Jensen}}, \bibinfo
  {author} {\bibfnamefont {S.}~\bibnamefont {Chemerisov}}, \ and\ \bibinfo
  {author} {\bibfnamefont {D.}~\bibnamefont {Budker}},\ }\href {\doibase
  10.1103/PhysRevLett.108.197601} {\bibfield  {journal} {\bibinfo  {journal}
  {Phys. Rev. Lett.}\ }\textbf {\bibinfo {volume} {108}},\ \bibinfo {pages}
  {197601} (\bibinfo {year} {2012})}\BibitemShut {NoStop}%
\bibitem [{\citenamefont {Grezes}\ \emph {et~al.}(2014)\citenamefont {Grezes},
  \citenamefont {Julsgaard}, \citenamefont {Kubo}, \citenamefont {Stern},
  \citenamefont {Umeda}, \citenamefont {Isoya}, \citenamefont {Sumiya},
  \citenamefont {Abe}, \citenamefont {Onoda}, \citenamefont {Ohshima},
  \citenamefont {Jacques}, \citenamefont {Esteve}, \citenamefont {Vion},
  \citenamefont {Esteve}, \citenamefont {Mølmer},\ and\ \citenamefont
  {Bertet}}]{grezes_multimode_2014}%
  \BibitemOpen
  \bibfield  {author} {\bibinfo {author} {\bibfnamefont {C.}~\bibnamefont
  {Grezes}}, \bibinfo {author} {\bibfnamefont {B.}~\bibnamefont {Julsgaard}},
  \bibinfo {author} {\bibfnamefont {Y.}~\bibnamefont {Kubo}}, \bibinfo {author}
  {\bibfnamefont {M.}~\bibnamefont {Stern}}, \bibinfo {author} {\bibfnamefont
  {T.}~\bibnamefont {Umeda}}, \bibinfo {author} {\bibfnamefont
  {J.}~\bibnamefont {Isoya}}, \bibinfo {author} {\bibfnamefont
  {H.}~\bibnamefont {Sumiya}}, \bibinfo {author} {\bibfnamefont
  {H.}~\bibnamefont {Abe}}, \bibinfo {author} {\bibfnamefont {S.}~\bibnamefont
  {Onoda}}, \bibinfo {author} {\bibfnamefont {T.}~\bibnamefont {Ohshima}},
  \bibinfo {author} {\bibfnamefont {V.}~\bibnamefont {Jacques}}, \bibinfo
  {author} {\bibfnamefont {J.}~\bibnamefont {Esteve}}, \bibinfo {author}
  {\bibfnamefont {D.}~\bibnamefont {Vion}}, \bibinfo {author} {\bibfnamefont
  {D.}~\bibnamefont {Esteve}}, \bibinfo {author} {\bibfnamefont
  {K.}~\bibnamefont {Mølmer}}, \ and\ \bibinfo {author} {\bibfnamefont
  {P.}~\bibnamefont {Bertet}},\ }\href {\doibase 10.1103/PhysRevX.4.021049}
  {\bibfield  {journal} {\bibinfo  {journal} {Physical Review X}\ }\textbf
  {\bibinfo {volume} {4}},\ \bibinfo {pages} {021049} (\bibinfo {year}
  {2014})}\BibitemShut {NoStop}%
\bibitem [{\citenamefont {Mizuochi}\ \emph {et~al.}(2009)\citenamefont
  {Mizuochi}, \citenamefont {Neumann}, \citenamefont {Rempp}, \citenamefont
  {Beck}, \citenamefont {Jacques}, \citenamefont {Siyushev}, \citenamefont
  {Nakamura}, \citenamefont {Twitchen}, \citenamefont {Watanabe}, \citenamefont
  {Yamasaki}, \citenamefont {Jelezko},\ and\ \citenamefont
  {Wrachtrup}}]{mizuochi_coherence_2009}%
  \BibitemOpen
  \bibfield  {author} {\bibinfo {author} {\bibfnamefont {N.}~\bibnamefont
  {Mizuochi}}, \bibinfo {author} {\bibfnamefont {P.}~\bibnamefont {Neumann}},
  \bibinfo {author} {\bibfnamefont {F.}~\bibnamefont {Rempp}}, \bibinfo
  {author} {\bibfnamefont {J.}~\bibnamefont {Beck}}, \bibinfo {author}
  {\bibfnamefont {V.}~\bibnamefont {Jacques}}, \bibinfo {author} {\bibfnamefont
  {P.}~\bibnamefont {Siyushev}}, \bibinfo {author} {\bibfnamefont
  {K.}~\bibnamefont {Nakamura}}, \bibinfo {author} {\bibfnamefont {D.~J.}\
  \bibnamefont {Twitchen}}, \bibinfo {author} {\bibfnamefont {H.}~\bibnamefont
  {Watanabe}}, \bibinfo {author} {\bibfnamefont {S.}~\bibnamefont {Yamasaki}},
  \bibinfo {author} {\bibfnamefont {F.}~\bibnamefont {Jelezko}}, \ and\
  \bibinfo {author} {\bibfnamefont {J.}~\bibnamefont {Wrachtrup}},\ }\href
  {\doibase 10.1103/PhysRevB.80.041201} {\bibfield  {journal} {\bibinfo
  {journal} {Phys. Rev. B}\ }\textbf {\bibinfo {volume} {80}},\ \bibinfo
  {pages} {041201} (\bibinfo {year} {2009})}\BibitemShut {NoStop}%
\bibitem [{\citenamefont {Maurer}\ \emph {et~al.}(2012)\citenamefont {Maurer},
  \citenamefont {Kucsko}, \citenamefont {Latta}, \citenamefont {Jiang},
  \citenamefont {Yao}, \citenamefont {Bennett}, \citenamefont {Pastawski},
  \citenamefont {Hunger}, \citenamefont {Chisholm}, \citenamefont {Markham},
  \citenamefont {Twitchen}, \citenamefont {Cirac},\ and\ \citenamefont
  {Lukin}}]{maurer_room-temperature_2012}%
  \BibitemOpen
  \bibfield  {author} {\bibinfo {author} {\bibfnamefont {P.~C.}\ \bibnamefont
  {Maurer}}, \bibinfo {author} {\bibfnamefont {G.}~\bibnamefont {Kucsko}},
  \bibinfo {author} {\bibfnamefont {C.}~\bibnamefont {Latta}}, \bibinfo
  {author} {\bibfnamefont {L.}~\bibnamefont {Jiang}}, \bibinfo {author}
  {\bibfnamefont {N.~Y.}\ \bibnamefont {Yao}}, \bibinfo {author} {\bibfnamefont
  {S.~D.}\ \bibnamefont {Bennett}}, \bibinfo {author} {\bibfnamefont
  {F.}~\bibnamefont {Pastawski}}, \bibinfo {author} {\bibfnamefont
  {D.}~\bibnamefont {Hunger}}, \bibinfo {author} {\bibfnamefont
  {N.}~\bibnamefont {Chisholm}}, \bibinfo {author} {\bibfnamefont
  {M.}~\bibnamefont {Markham}}, \bibinfo {author} {\bibfnamefont {D.~J.}\
  \bibnamefont {Twitchen}}, \bibinfo {author} {\bibfnamefont {J.~I.}\
  \bibnamefont {Cirac}}, \ and\ \bibinfo {author} {\bibfnamefont {M.~D.}\
  \bibnamefont {Lukin}},\ }\href {\doibase 10.1126/science.1220513} {\bibfield
  {journal} {\bibinfo  {journal} {Science}\ }\textbf {\bibinfo {volume}
  {336}},\ \bibinfo {pages} {1283} (\bibinfo {year} {2012})}\BibitemShut
  {NoStop}%
\bibitem [{\citenamefont {Balasubramian}\ \emph {et~al.}(2009)\citenamefont
  {Balasubramian}, \citenamefont {Neumann}, \citenamefont {Twitchen},
  \citenamefont {Markham}, \citenamefont {Koselov}, \citenamefont {Mizuochi},
  \citenamefont {Isoya}, \citenamefont {Achard}, \citenamefont {Beck},
  \citenamefont {Tissler}, \citenamefont {Jacques}, \citenamefont {Hemmer},
  \citenamefont {Jelezko},\ and\ \citenamefont
  {Wrachtrup}}]{Balasubramian.NatMat.8.383(2009)}%
  \BibitemOpen
  \bibfield  {author} {\bibinfo {author} {\bibfnamefont {G.}~\bibnamefont
  {Balasubramian}}, \bibinfo {author} {\bibfnamefont {P.}~\bibnamefont
  {Neumann}}, \bibinfo {author} {\bibfnamefont {D.}~\bibnamefont {Twitchen}},
  \bibinfo {author} {\bibfnamefont {M.}~\bibnamefont {Markham}}, \bibinfo
  {author} {\bibfnamefont {R.}~\bibnamefont {Koselov}}, \bibinfo {author}
  {\bibfnamefont {N.}~\bibnamefont {Mizuochi}}, \bibinfo {author}
  {\bibfnamefont {J.}~\bibnamefont {Isoya}}, \bibinfo {author} {\bibfnamefont
  {J.}~\bibnamefont {Achard}}, \bibinfo {author} {\bibfnamefont
  {J.}~\bibnamefont {Beck}}, \bibinfo {author} {\bibfnamefont {J.}~\bibnamefont
  {Tissler}}, \bibinfo {author} {\bibfnamefont {V.}~\bibnamefont {Jacques}},
  \bibinfo {author} {\bibfnamefont {P.}~\bibnamefont {Hemmer}}, \bibinfo
  {author} {\bibfnamefont {F.}~\bibnamefont {Jelezko}}, \ and\ \bibinfo
  {author} {\bibfnamefont {J.}~\bibnamefont {Wrachtrup}},\ }\href {\doibase
  10.1038/nmat2420} {\bibfield  {journal} {\bibinfo  {journal} {Nature
  Materials}\ }\textbf {\bibinfo {volume} {8}},\ \bibinfo {pages} {383}
  (\bibinfo {year} {2009})}\BibitemShut {NoStop}%
\bibitem [{\citenamefont {Yamamoto}\ \emph
  {et~al.}(2013{\natexlab{a}})\citenamefont {Yamamoto}, \citenamefont {Umeda},
  \citenamefont {Watanabe}, \citenamefont {Onoda}, \citenamefont {Markham},
  \citenamefont {Twitchen}, \citenamefont {Naydenov}, \citenamefont
  {McGuinness}, \citenamefont {Teraji}, \citenamefont {Koizumi}, \citenamefont
  {Dolde}, \citenamefont {Fedder}, \citenamefont {Honert}, \citenamefont
  {Wrachtrup}, \citenamefont {Ohshima}, \citenamefont {Jelezko},\ and\
  \citenamefont {Isoya}}]{yamamoto_extending_2013}%
  \BibitemOpen
  \bibfield  {author} {\bibinfo {author} {\bibfnamefont {T.}~\bibnamefont
  {Yamamoto}}, \bibinfo {author} {\bibfnamefont {T.}~\bibnamefont {Umeda}},
  \bibinfo {author} {\bibfnamefont {K.}~\bibnamefont {Watanabe}}, \bibinfo
  {author} {\bibfnamefont {S.}~\bibnamefont {Onoda}}, \bibinfo {author}
  {\bibfnamefont {M.~L.}\ \bibnamefont {Markham}}, \bibinfo {author}
  {\bibfnamefont {D.~J.}\ \bibnamefont {Twitchen}}, \bibinfo {author}
  {\bibfnamefont {B.}~\bibnamefont {Naydenov}}, \bibinfo {author}
  {\bibfnamefont {L.~P.}\ \bibnamefont {McGuinness}}, \bibinfo {author}
  {\bibfnamefont {T.}~\bibnamefont {Teraji}}, \bibinfo {author} {\bibfnamefont
  {S.}~\bibnamefont {Koizumi}}, \bibinfo {author} {\bibfnamefont
  {F.}~\bibnamefont {Dolde}}, \bibinfo {author} {\bibfnamefont
  {H.}~\bibnamefont {Fedder}}, \bibinfo {author} {\bibfnamefont
  {J.}~\bibnamefont {Honert}}, \bibinfo {author} {\bibfnamefont
  {J.}~\bibnamefont {Wrachtrup}}, \bibinfo {author} {\bibfnamefont
  {T.}~\bibnamefont {Ohshima}}, \bibinfo {author} {\bibfnamefont
  {F.}~\bibnamefont {Jelezko}}, \ and\ \bibinfo {author} {\bibfnamefont
  {J.}~\bibnamefont {Isoya}},\ }\href {\doibase 10.1103/PhysRevB.88.075206}
  {\bibfield  {journal} {\bibinfo  {journal} {Phys. Rev. B}\ }\textbf {\bibinfo
  {volume} {88}},\ \bibinfo {pages} {075206} (\bibinfo {year}
  {2013}{\natexlab{a}})}\BibitemShut {NoStop}%
\bibitem [{\citenamefont {Yamamoto}\ \emph
  {et~al.}(2013{\natexlab{b}})\citenamefont {Yamamoto}, \citenamefont
  {M\"uller}, \citenamefont {McGuinness}, \citenamefont {Teraji}, \citenamefont
  {Naydenov}, \citenamefont {Onoda}, \citenamefont {Ohshima}, \citenamefont
  {Wrachtrup}, \citenamefont {Jelezko},\ and\ \citenamefont
  {Isoya}}]{yamamoto_strongly_2013}%
  \BibitemOpen
  \bibfield  {author} {\bibinfo {author} {\bibfnamefont {T.}~\bibnamefont
  {Yamamoto}}, \bibinfo {author} {\bibfnamefont {C.}~\bibnamefont {M\"uller}},
  \bibinfo {author} {\bibfnamefont {L.~P.}\ \bibnamefont {McGuinness}},
  \bibinfo {author} {\bibfnamefont {T.}~\bibnamefont {Teraji}}, \bibinfo
  {author} {\bibfnamefont {B.}~\bibnamefont {Naydenov}}, \bibinfo {author}
  {\bibfnamefont {S.}~\bibnamefont {Onoda}}, \bibinfo {author} {\bibfnamefont
  {T.}~\bibnamefont {Ohshima}}, \bibinfo {author} {\bibfnamefont
  {J.}~\bibnamefont {Wrachtrup}}, \bibinfo {author} {\bibfnamefont
  {F.}~\bibnamefont {Jelezko}}, \ and\ \bibinfo {author} {\bibfnamefont
  {J.}~\bibnamefont {Isoya}},\ }\href {\doibase 10.1103/PhysRevB.88.201201}
  {\bibfield  {journal} {\bibinfo  {journal} {Phys. Rev. B}\ }\textbf {\bibinfo
  {volume} {88}},\ \bibinfo {pages} {201201} (\bibinfo {year}
  {2013}{\natexlab{b}})}\BibitemShut {NoStop}%
\bibitem [{\citenamefont {Feher}\ and\ \citenamefont
  {Gere}(1959)}]{Feher1959a}%
  \BibitemOpen
  \bibfield  {author} {\bibinfo {author} {\bibfnamefont {G.}~\bibnamefont
  {Feher}}\ and\ \bibinfo {author} {\bibfnamefont {E.}~\bibnamefont {Gere}},\
  }\href {\doibase 10.1103/PhysRev.114.1245} {\enquote {\bibinfo {title}
  {{Electron Spin Resonance Experiments on Donors in Silicon. II. Electron Spin
  Relaxation Effects}},}\ } (\bibinfo {year} {1959})\BibitemShut {NoStop}%
\bibitem [{\citenamefont {Morley}\ \emph {et~al.}(2010)\citenamefont {Morley},
  \citenamefont {Warner}, \citenamefont {Stoneham}, \citenamefont {Greenland},
  \citenamefont {van Tol}, \citenamefont {Kay},\ and\ \citenamefont
  {Aeppli}}]{Morley.NatureMat.9.725(2010)}%
  \BibitemOpen
  \bibfield  {author} {\bibinfo {author} {\bibfnamefont {G.~W.}\ \bibnamefont
  {Morley}}, \bibinfo {author} {\bibfnamefont {M.}~\bibnamefont {Warner}},
  \bibinfo {author} {\bibfnamefont {A.~M.}\ \bibnamefont {Stoneham}}, \bibinfo
  {author} {\bibfnamefont {P.~T.}\ \bibnamefont {Greenland}}, \bibinfo {author}
  {\bibfnamefont {J.}~\bibnamefont {van Tol}}, \bibinfo {author} {\bibfnamefont
  {C.~W.}\ \bibnamefont {Kay}}, \ and\ \bibinfo {author} {\bibfnamefont
  {G.}~\bibnamefont {Aeppli}},\ }\href@noop {} {\bibfield  {journal} {\bibinfo
  {journal} {Nature materials}\ }\textbf {\bibinfo {volume} {9}},\ \bibinfo
  {pages} {725} (\bibinfo {year} {2010})}\BibitemShut {NoStop}%
\bibitem [{\citenamefont {Wolfowicz}\ \emph {et~al.}(2012)\citenamefont
  {Wolfowicz}, \citenamefont {Simmons}, \citenamefont {Tyryshkin},
  \citenamefont {George}, \citenamefont {Riemann}, \citenamefont {Abrosimov},
  \citenamefont {Becker}, \citenamefont {Pohl}, \citenamefont {Lyon},
  \citenamefont {Thewalt},\ and\ \citenamefont
  {Morton}}]{wolfowicz_decoherence_2012}%
  \BibitemOpen
  \bibfield  {author} {\bibinfo {author} {\bibfnamefont {G.}~\bibnamefont
  {Wolfowicz}}, \bibinfo {author} {\bibfnamefont {S.}~\bibnamefont {Simmons}},
  \bibinfo {author} {\bibfnamefont {A.~M.}\ \bibnamefont {Tyryshkin}}, \bibinfo
  {author} {\bibfnamefont {R.~E.}\ \bibnamefont {George}}, \bibinfo {author}
  {\bibfnamefont {H.}~\bibnamefont {Riemann}}, \bibinfo {author} {\bibfnamefont
  {N.~V.}\ \bibnamefont {Abrosimov}}, \bibinfo {author} {\bibfnamefont
  {P.}~\bibnamefont {Becker}}, \bibinfo {author} {\bibfnamefont {H.-J.}\
  \bibnamefont {Pohl}}, \bibinfo {author} {\bibfnamefont {S.~A.}\ \bibnamefont
  {Lyon}}, \bibinfo {author} {\bibfnamefont {M.~L.~W.}\ \bibnamefont
  {Thewalt}}, \ and\ \bibinfo {author} {\bibfnamefont {J.~J.~L.}\ \bibnamefont
  {Morton}},\ }\href {\doibase 10.1103/PhysRevB.86.245301} {\bibfield
  {journal} {\bibinfo  {journal} {Phys. Rev. B}\ }\textbf {\bibinfo {volume}
  {86}},\ \bibinfo {pages} {245301} (\bibinfo {year} {2012})}\BibitemShut
  {NoStop}%
\bibitem [{\citenamefont {Tyryshkin}\ \emph {et~al.}(2011)\citenamefont
  {Tyryshkin}, \citenamefont {Tojo}, \citenamefont {Morton}, \citenamefont
  {Riemann}, \citenamefont {Abrosimov}, \citenamefont {Becker}, \citenamefont
  {Pohl}, \citenamefont {Schenkel}, \citenamefont {Thewalt}, \citenamefont
  {Itoh},\ and\ \citenamefont {Lyon}}]{tyryshkin_electron_2011}%
  \BibitemOpen
  \bibfield  {author} {\bibinfo {author} {\bibfnamefont {A.~M.}\ \bibnamefont
  {Tyryshkin}}, \bibinfo {author} {\bibfnamefont {S.}~\bibnamefont {Tojo}},
  \bibinfo {author} {\bibfnamefont {J.~J.~L.}\ \bibnamefont {Morton}}, \bibinfo
  {author} {\bibfnamefont {H.}~\bibnamefont {Riemann}}, \bibinfo {author}
  {\bibfnamefont {N.~V.}\ \bibnamefont {Abrosimov}}, \bibinfo {author}
  {\bibfnamefont {P.}~\bibnamefont {Becker}}, \bibinfo {author} {\bibfnamefont
  {H.-J.}\ \bibnamefont {Pohl}}, \bibinfo {author} {\bibfnamefont
  {T.}~\bibnamefont {Schenkel}}, \bibinfo {author} {\bibfnamefont {M.~L.~W.}\
  \bibnamefont {Thewalt}}, \bibinfo {author} {\bibfnamefont {K.~M.}\
  \bibnamefont {Itoh}}, \ and\ \bibinfo {author} {\bibfnamefont {S.~A.}\
  \bibnamefont {Lyon}},\ }\href {\doibase 10.1038/nmat3182} {\bibfield
  {journal} {\bibinfo  {journal} {Nature Materials}\ }\textbf {\bibinfo
  {volume} {11}},\ \bibinfo {pages} {143} (\bibinfo {year} {2011})}\BibitemShut
  {NoStop}%
\bibitem [{\citenamefont {Weis}\ \emph {et~al.}(2012)\citenamefont {Weis},
  \citenamefont {Lo}, \citenamefont {Lang}, \citenamefont {Tyryshkin},
  \citenamefont {George}, \citenamefont {Yu}, \citenamefont {Bokor},
  \citenamefont {Lyon}, \citenamefont {Morton},\ and\ \citenamefont
  {Schenkel}}]{Weis.APL.100.172104(2012)}%
  \BibitemOpen
  \bibfield  {author} {\bibinfo {author} {\bibfnamefont {C.~D.}\ \bibnamefont
  {Weis}}, \bibinfo {author} {\bibfnamefont {C.~C.}\ \bibnamefont {Lo}},
  \bibinfo {author} {\bibfnamefont {V.}~\bibnamefont {Lang}}, \bibinfo {author}
  {\bibfnamefont {A.~M.}\ \bibnamefont {Tyryshkin}}, \bibinfo {author}
  {\bibfnamefont {R.~E.}\ \bibnamefont {George}}, \bibinfo {author}
  {\bibfnamefont {K.~M.}\ \bibnamefont {Yu}}, \bibinfo {author} {\bibfnamefont
  {J.}~\bibnamefont {Bokor}}, \bibinfo {author} {\bibfnamefont {S.~A.}\
  \bibnamefont {Lyon}}, \bibinfo {author} {\bibfnamefont {J.~J.~L.}\
  \bibnamefont {Morton}}, \ and\ \bibinfo {author} {\bibfnamefont
  {T.}~\bibnamefont {Schenkel}},\ }\href {\doibase
  http://dx.doi.org/10.1063/1.4704561} {\bibfield  {journal} {\bibinfo
  {journal} {Applied Physics Letters}\ }\textbf {\bibinfo {volume} {100}},\
  \bibinfo {eid} {172104} (\bibinfo {year} {2012})}\BibitemShut {NoStop}%
\bibitem [{\citenamefont {Muhonen}\ \emph {et~al.}(2014)\citenamefont
  {Muhonen}, \citenamefont {Dehollain}, \citenamefont {Laucht}, \citenamefont
  {Hudson}, \citenamefont {Kalra}, \citenamefont {Sekiguchi}, \citenamefont
  {Itoh}, \citenamefont {Jamieson}, \citenamefont {McCallum}, \citenamefont
  {Dzurak},\ and\ \citenamefont {Morello}}]{muhonen_storing_2014}%
  \BibitemOpen
  \bibfield  {author} {\bibinfo {author} {\bibfnamefont {J.~T.}\ \bibnamefont
  {Muhonen}}, \bibinfo {author} {\bibfnamefont {J.~P.}\ \bibnamefont
  {Dehollain}}, \bibinfo {author} {\bibfnamefont {A.}~\bibnamefont {Laucht}},
  \bibinfo {author} {\bibfnamefont {F.~E.}\ \bibnamefont {Hudson}}, \bibinfo
  {author} {\bibfnamefont {R.}~\bibnamefont {Kalra}}, \bibinfo {author}
  {\bibfnamefont {T.}~\bibnamefont {Sekiguchi}}, \bibinfo {author}
  {\bibfnamefont {K.~M.}\ \bibnamefont {Itoh}}, \bibinfo {author}
  {\bibfnamefont {D.~N.}\ \bibnamefont {Jamieson}}, \bibinfo {author}
  {\bibfnamefont {J.~C.}\ \bibnamefont {McCallum}}, \bibinfo {author}
  {\bibfnamefont {A.~S.}\ \bibnamefont {Dzurak}}, \ and\ \bibinfo {author}
  {\bibfnamefont {A.}~\bibnamefont {Morello}},\ }\href {\doibase
  10.1038/nnano.2014.211} {\bibfield  {journal} {\bibinfo  {journal} {Nature
  Nanotechnology}\ }\textbf {\bibinfo {volume} {9}},\ \bibinfo {pages} {986}
  (\bibinfo {year} {2014})}\BibitemShut {NoStop}%
\bibitem [{\citenamefont {Yurke}\ and\ \citenamefont
  {Denker}(1984)}]{Yurke(1984)}%
  \BibitemOpen
  \bibfield  {author} {\bibinfo {author} {\bibfnamefont {B.}~\bibnamefont
  {Yurke}}\ and\ \bibinfo {author} {\bibfnamefont {J.~S.}\ \bibnamefont
  {Denker}},\ }\href {\doibase 10.1103/PhysRevA.29.1419} {\bibfield  {journal}
  {\bibinfo  {journal} {Phys. Rev. A}\ }\textbf {\bibinfo {volume} {29}},\
  \bibinfo {pages} {1419} (\bibinfo {year} {1984})}\BibitemShut {NoStop}%
\bibitem [{\citenamefont {Zgirski}\ \emph {et~al.}(2005)\citenamefont
  {Zgirski}, \citenamefont {Riikonen}, \citenamefont {Touboltsev},\ and\
  \citenamefont {Arutyunov}}]{Zgirski.NanoLett.5.1029(2005)}%
  \BibitemOpen
  \bibfield  {author} {\bibinfo {author} {\bibfnamefont {M.}~\bibnamefont
  {Zgirski}}, \bibinfo {author} {\bibfnamefont {K.-P.}\ \bibnamefont
  {Riikonen}}, \bibinfo {author} {\bibfnamefont {V.}~\bibnamefont
  {Touboltsev}}, \ and\ \bibinfo {author} {\bibfnamefont {K.}~\bibnamefont
  {Arutyunov}},\ }\href@noop {} {\bibfield  {journal} {\bibinfo  {journal}
  {Nano letters}\ }\textbf {\bibinfo {volume} {5}},\ \bibinfo {pages} {1029}
  (\bibinfo {year} {2005})}\BibitemShut {NoStop}%
\bibitem [{\citenamefont {Megrant}\ \emph {et~al.}(2012)\citenamefont
  {Megrant}, \citenamefont {Neill}, \citenamefont {Barends}, \citenamefont
  {Chiaro}, \citenamefont {Chen}, \citenamefont {Feigl}, \citenamefont {Kelly},
  \citenamefont {Lucero}, \citenamefont {Mariantoni}, \citenamefont {O'Malley},
  \citenamefont {Sank}, \citenamefont {Vainsencher}, \citenamefont {Wenner},
  \citenamefont {White}, \citenamefont {Yin}, \citenamefont {Zhao},
  \citenamefont {Palmstr�m}, \citenamefont {Martinis},\ and\ \citenamefont
  {Cleland}}]{Megrant.APL.100.113510(2012)}%
  \BibitemOpen
  \bibfield  {author} {\bibinfo {author} {\bibfnamefont {A.}~\bibnamefont
  {Megrant}}, \bibinfo {author} {\bibfnamefont {C.}~\bibnamefont {Neill}},
  \bibinfo {author} {\bibfnamefont {R.}~\bibnamefont {Barends}}, \bibinfo
  {author} {\bibfnamefont {B.}~\bibnamefont {Chiaro}}, \bibinfo {author}
  {\bibfnamefont {Y.}~\bibnamefont {Chen}}, \bibinfo {author} {\bibfnamefont
  {L.}~\bibnamefont {Feigl}}, \bibinfo {author} {\bibfnamefont
  {J.}~\bibnamefont {Kelly}}, \bibinfo {author} {\bibfnamefont
  {E.}~\bibnamefont {Lucero}}, \bibinfo {author} {\bibfnamefont
  {M.}~\bibnamefont {Mariantoni}}, \bibinfo {author} {\bibfnamefont {P.~J.~J.}\
  \bibnamefont {O'Malley}}, \bibinfo {author} {\bibfnamefont {D.}~\bibnamefont
  {Sank}}, \bibinfo {author} {\bibfnamefont {A.}~\bibnamefont {Vainsencher}},
  \bibinfo {author} {\bibfnamefont {J.}~\bibnamefont {Wenner}}, \bibinfo
  {author} {\bibfnamefont {T.~C.}\ \bibnamefont {White}}, \bibinfo {author}
  {\bibfnamefont {Y.}~\bibnamefont {Yin}}, \bibinfo {author} {\bibfnamefont
  {J.}~\bibnamefont {Zhao}}, \bibinfo {author} {\bibfnamefont {C.~J.}\
  \bibnamefont {Palmstr�m}}, \bibinfo {author} {\bibfnamefont {J.~M.}\
  \bibnamefont {Martinis}}, \ and\ \bibinfo {author} {\bibfnamefont {A.~N.}\
  \bibnamefont {Cleland}},\ }\href@noop {} {\bibfield  {journal} {\bibinfo
  {journal} {Appl. Phys. Lett.}\ }\textbf {\bibinfo {volume} {100}},\ \bibinfo
  {pages} {113510} (\bibinfo {year} {2012})}\BibitemShut {NoStop}%
\bibitem [{\citenamefont {Annunziata}\ \emph {et~al.}(2010)\citenamefont
  {Annunziata}, \citenamefont {Santavicca}, \citenamefont {Frunzio},
  \citenamefont {Catelani}, \citenamefont {Rooks}, \citenamefont {Frydman},\
  and\ \citenamefont {Prober}}]{Lk_formula}%
  \BibitemOpen
  \bibfield  {author} {\bibinfo {author} {\bibfnamefont {A.~J.}\ \bibnamefont
  {Annunziata}}, \bibinfo {author} {\bibfnamefont {D.~F.}\ \bibnamefont
  {Santavicca}}, \bibinfo {author} {\bibfnamefont {L.}~\bibnamefont {Frunzio}},
  \bibinfo {author} {\bibfnamefont {G.}~\bibnamefont {Catelani}}, \bibinfo
  {author} {\bibfnamefont {M.~J.}\ \bibnamefont {Rooks}}, \bibinfo {author}
  {\bibfnamefont {A.}~\bibnamefont {Frydman}}, \ and\ \bibinfo {author}
  {\bibfnamefont {D.~E.}\ \bibnamefont {Prober}},\ }\href
  {http://stacks.iop.org/0957-4484/21/i=44/a=445202} {\bibfield  {journal}
  {\bibinfo  {journal} {Nanotechnology}\ }\textbf {\bibinfo {volume} {21}},\
  \bibinfo {pages} {445202} (\bibinfo {year} {2010})}\BibitemShut {NoStop}%
\bibitem [{\citenamefont {Vries}(1988)}]{ThinFilm_Resistivity_DeVries1988}%
  \BibitemOpen
  \bibfield  {author} {\bibinfo {author} {\bibfnamefont {J.~D.}\ \bibnamefont
  {Vries}},\ }\href {\doibase http://dx.doi.org/10.1016/0040-6090(88)90478-6}
  {\bibfield  {journal} {\bibinfo  {journal} {Thin Solid Films}\ }\textbf
  {\bibinfo {volume} {167}},\ \bibinfo {pages} {25 } (\bibinfo {year}
  {1988})}\BibitemShut {NoStop}%
\bibitem [{\citenamefont {Court}\ \emph {et~al.}(2008)\citenamefont {Court},
  \citenamefont {Ferguson},\ and\ \citenamefont {Clark}}]{AlThinFilmGap}%
  \BibitemOpen
  \bibfield  {author} {\bibinfo {author} {\bibfnamefont {N.~A.}\ \bibnamefont
  {Court}}, \bibinfo {author} {\bibfnamefont {A.~J.}\ \bibnamefont {Ferguson}},
  \ and\ \bibinfo {author} {\bibfnamefont {R.~G.}\ \bibnamefont {Clark}},\
  }\href {http://stacks.iop.org/0953-2048/21/i=1/a=015013} {\bibfield
  {journal} {\bibinfo  {journal} {Superconductor Science and Technology}\
  }\textbf {\bibinfo {volume} {21}},\ \bibinfo {pages} {015013} (\bibinfo
  {year} {2008})}\BibitemShut {NoStop}%
\bibitem [{\citenamefont {Julsgaard}\ and\ \citenamefont
  {M\o{}lmer}(2012)}]{julsgaard_measurement-induced_2012}%
  \BibitemOpen
  \bibfield  {author} {\bibinfo {author} {\bibfnamefont {B.}~\bibnamefont
  {Julsgaard}}\ and\ \bibinfo {author} {\bibfnamefont {K.}~\bibnamefont
  {M\o{}lmer}},\ }\href {\doibase 10.1103/PhysRevA.85.032327} {\bibfield
  {journal} {\bibinfo  {journal} {Phys. Rev. A}\ }\textbf {\bibinfo {volume}
  {85}},\ \bibinfo {pages} {032327} (\bibinfo {year} {2012})}\BibitemShut
  {NoStop}%
\bibitem [{\citenamefont {Bienfait}\ \emph
  {et~al.}(2016{\natexlab{b}})\citenamefont {Bienfait}, \citenamefont {Pla},
  \citenamefont {Kubo}, \citenamefont {Stern}, \citenamefont {Zhou},
  \citenamefont {Lo}, \citenamefont {Weis}, \citenamefont {Schenkel},
  \citenamefont {Vion}, \citenamefont {Esteve}, \citenamefont {Morton},\ and\
  \citenamefont {Bertet}}]{bienfait2015controlling}%
  \BibitemOpen
  \bibfield  {author} {\bibinfo {author} {\bibfnamefont {A.}~\bibnamefont
  {Bienfait}}, \bibinfo {author} {\bibfnamefont {J.}~\bibnamefont {Pla}},
  \bibinfo {author} {\bibfnamefont {Y.}~\bibnamefont {Kubo}}, \bibinfo {author}
  {\bibfnamefont {M.}~\bibnamefont {Stern}}, \bibinfo {author} {\bibfnamefont
  {X.}~\bibnamefont {Zhou}}, \bibinfo {author} {\bibfnamefont {C.}~\bibnamefont
  {Lo}}, \bibinfo {author} {\bibfnamefont {C.}~\bibnamefont {Weis}}, \bibinfo
  {author} {\bibfnamefont {T.}~\bibnamefont {Schenkel}}, \bibinfo {author}
  {\bibfnamefont {D.}~\bibnamefont {Vion}}, \bibinfo {author} {\bibfnamefont
  {D.}~\bibnamefont {Esteve}}, \bibinfo {author} {\bibfnamefont
  {J.}~\bibnamefont {Morton}}, \ and\ \bibinfo {author} {\bibfnamefont
  {P.}~\bibnamefont {Bertet}},\ }\href@noop {} {\bibfield  {journal} {\bibinfo
  {journal} {Nature}\ }\textbf {\bibinfo {volume} {531}},\ \bibinfo {pages}
  {74} (\bibinfo {year} {2016}{\natexlab{b}})}\BibitemShut {NoStop}%
\bibitem [{\citenamefont {Kiilerich}\ and\ \citenamefont
  {M\o{}lmer}(2014)}]{Kiilerich.PhysRevA.89.052110(2014)}%
  \BibitemOpen
  \bibfield  {author} {\bibinfo {author} {\bibfnamefont {A.~H.}\ \bibnamefont
  {Kiilerich}}\ and\ \bibinfo {author} {\bibfnamefont {K.}~\bibnamefont
  {M\o{}lmer}},\ }\href {\doibase 10.1103/PhysRevA.89.052110} {\bibfield
  {journal} {\bibinfo  {journal} {Phys. Rev. A}\ }\textbf {\bibinfo {volume}
  {89}},\ \bibinfo {pages} {052110} (\bibinfo {year} {2014})}\BibitemShut
  {NoStop}%
\bibitem [{\citenamefont {Kiilerich}\ and\ \citenamefont
  {M\o{}lmer}(2015)}]{Kiilerich.PhysRevA.91.012119(2015)}%
  \BibitemOpen
  \bibfield  {author} {\bibinfo {author} {\bibfnamefont {A.~H.}\ \bibnamefont
  {Kiilerich}}\ and\ \bibinfo {author} {\bibfnamefont {K.}~\bibnamefont
  {M\o{}lmer}},\ }\href {\doibase 10.1103/PhysRevA.91.012119} {\bibfield
  {journal} {\bibinfo  {journal} {Phys. Rev. A}\ }\textbf {\bibinfo {volume}
  {91}},\ \bibinfo {pages} {012119} (\bibinfo {year} {2015})}\BibitemShut
  {NoStop}%
\end{thebibliography}
\end{document}